\documentclass[12pt,preprint]{aastex} 
\usepackage{emulateapj5}

\newcommand{\be}{\begin{equation}}
\newcommand{\ee}{\end{equation}}
\newcommand{\beqn}{\begin{eqnarray}}
\newcommand{\eeqn}{\end{eqnarray}}
\newcommand{\hu} {{{\bld{u}_{J_x,J_y,J_z}}}}
\newcommand{\bld}[1]{\mbox{\boldmath$#1$\unboldmath}}

\newcommand{\bky}{{\bar{K}_y}}

\begin{document}

\title{Nonlinear Evolution of Hydrodynamical Shear Flows in Two Dimensions}

\author{Yoram Lithwick\altaffilmark{1}}
\altaffiltext{1}{CITA. Toronto, Ontario, Canada; yoram@cita.utoronto.ca}

\begin{abstract}
We examine how perturbed shear flows evolve in two-dimensional, incompressible, inviscid
hydrodynamical
fluids,
with the ultimate goal of understanding the dynamics of accretion disks, as well as other
astrophysical shear flows.
To linear order,  vorticity waves are swung around by the background shear, and their velocities
are amplified transiently before decaying.
It has been speculated that sufficiently amplified modes might couple nonlinearly, 
leading to turbulence.  Here we show how nonlinear coupling occurs in two dimensions.  
This coupling
 is remarkably simple because it only lasts for a short time interval, when one of the coupled modes
 is in mid-swing,
  i.e., when its phasefronts are aligned with the radial direction. We focus on the interaction between an
  axisymmetric mode and a  swinging mode.
  We find that all axisymmetric modes, regardless of how small in amplitude, are unstable when
  they interact with swinging modes that have sufficiently large azimuthal wavelength.
 Quantitatively, the criterion for instability is that
 $|k_{y,\rm sw}/k_{x,\rm axi}|\lesssim |\omega/q| $, i.e., that the ratio of wavenumbers
 (swinging azimuthal 
 wavenumber to axisymmetric radial wavenumber) is less than the ratio of the perturbed vorticity to 
 the background vorticity. If  this is the case, then when the swinging 
 mode is in mid-swing it couples with the axisymmetric mode to produce a new leading swinging mode 
 that has larger vorticity than itself; this new mode in turn produces an even larger leading mode, 
 etc.
  We explain how  this shear (or Kelvin-Helmholtz) instability operates in 
  real-space as well.
   The instability  occurs whenever the 
 momentum transported by 
 an energy conserving
  perturbation has the sign required for it to diminish the background shear; only when 
 this occurs can energy be extracted from the mean flow and hence added to the perturbation.
For an accretion disk, this means that the instability transports angular momentum outwards while
it operates.
We verify all our conclusions in detail with full hydrodynamical simulations, done with a pseudospectral method in a shearing box.  Simulations of the  instability form  vortices whose boundaries become highly
convoluted.
Whether this nonlinear instability plays a role
in accretion disks is an interesting possibility.

 \end{abstract}

\section{Introduction}

Matter accretes onto a wide variety of objects, such as young stars,
black holes, and white dwarfs,  through accretion disks.
For matter in an accretion disk
 to fall inwards, angular momentum must be transported outwards.
In many accretion disks, it is turbulence and the resulting
turbulent viscosity that is responsible for angular momentum
transport.
Therefore, if one wishes to understand accretion disks, one must 
first address
how disks 
become turbulent, and how much turbulence they sustain. 
In highly ionized disks, magnetic fields can trigger turbulence 
via the magnetorotational
instability \citep{BH98}.
However in neutral (hydrodynamical) disks, the situation is unclear.
 It remains an open question whether neutral Keplerian disks 
are turbulent, or whether they can remain laminar despite enormous Reynolds numbers.
Until it is answered, the accretion of nearly neutral disks, such as those
around young stars \citep[e.g., ][]{SMUN00} or dwarf novae \citep[e.g.,][]{GM98}, 
will remain mysteries.  
Simulations \citep{HBW99,SSG06} and experiments \citep{JBSG06}
both point to the answer that incompressible 
hydro disks are laminar---or at least
that any turbulence in them would not be strong enough to act 
as the agent  of angular momentum transport 
\citep{LL05}.
But the evidence is not conclusive because
the Reynolds numbers in disks  are much larger than
those accessible to computers or experiments.

Even if one takes the view that numerical and experimental evidence rules out
self-sustaining incompressible turbulence in Keplerian disks, it remains
important to understand how such shear flows work.  
In our view, this is an essential first step before more complicated effects---such
as baroclinic effects, convection, stirring by planets, stirring by dust, or
stirring by a magnetically active layer above the midplane---can be understood.
It could be that one of these effects is responsible for turbulence in disks.
In the present paper, we describe in detail
 the
dynamics of perturbations in incompressible shear flows in two dimensions (in the plane
of the disk).
Even though 2D flows do not lead to turbulence, they exhibit a remarkably rich dynamics.
Of course, incompressible 2D shear flows have been the subject of an enormous amount of research 
\citep[e.g., ][]{DR04}.
But our methods are different: we calculate the nonlinear coupling between linear waves.
One of the benefits of this approach is that it is straightforward to generalize it to
 three dimensions, as well as to include effects such compressibility and magnetohydrodynamics.

Recently, the evolution of linear waves has been the focus of much attention in the astrophysical
literature.  Since the velocity field of a linear swinging wave
can be amplified by very large factors during its swing,
 it has been proposed that sufficiently amplified modes might
couple nonlinearly, leading ultimately to turbulence \citep[e.g.,][]{IK01,CZTL03,UR04,Yecko04,AMN05}.  
However, how this coupling might occur has not been discussed (though see \citeauthor{Muk06} 
\citeyear{Muk06}),
 nor has the 
amplitude needed to trigger nonlinear effects been quantified.  The present paper begins to rectify these
shortcomings.
\cite{BH06} show that linear swinging waves are exact solutions of the nonlinear equations
of motion.  They contend that this argues against transient amplification as a route to turbulence
in unmagnetized disks.  However, we show in the present paper that nonlinear coupling
between two different waves can lead to interesting dynamics even when the individual waves
are exact nonlinear solutions.

Numerical simulations of 2D shear flows  starting from random initial conditions
settle into a distinctive banded structure 
\citep{UR04,JG05,SSG06}.  Roughly speaking, bands where $\omega>0$ are interspersed
with bands where $\omega<0$ ($\omega$ is the perturbed vorticity, eq. [\ref{eq:om}]).
Bands where $\omega$ has the same sign as the background vorticity
contain a single vortex;  in  bands with the opposite sign,  $\omega$ is smooth and there are no
vortices.
One of the goals of this paper is to explain why this is a natural outcome of random initial conditions.

\subsection{Organization}
We introduce the equations of motion in 
\S \ref{sec:eom}.  In  \S\S \ref{sec:fourier}-\ref{sec:ssw}, the heart of this paper,
 we describe and simulate the nonlinear
coupling between modes, focusing on the shear instability that results from the 
coupling between swinging waves and
axisymmetric ones. 
In \S \ref{sec:piece}, which can be read independently of
\S\S \ref{sec:fourier}-\ref{sec:ssw},
we describe the instability in real space.
We give both a dynamical explanation and one based on momentum and energy arguments.
We also simulate the fully nonlinear outcome of the instability.  We conclude in \S \ref{sec:summary}.
 In the Appendix, we describe the pseudospectral code that we use to run simulations
 in \S\S \ref{sec:fourier}-\ref{sec:ssw}.

\section{Equations of Motion}
\label{sec:eom}

An unperturbed Keplerian disk has angular velocity profile $\Omega(r)\propto r^{-3/2}$.
We write the fluid equations  in a reference
frame rotating at constant angular speed $\Omega_0\equiv \Omega(r_0)$, where $r_0$ is
a fiducial radius, and replace the radial and azimuthal coordinates  $r,\theta$  with Cartesian
coordinates, $x\equiv r-r_0$, $y\equiv r_0\theta$.  On lengthscales much smaller than $r_0$, the 
``shearing box'' equations of motion for an incompressible fluid are
\begin{eqnarray}
{\partial_t {\bld v}}+{\bld{ v\cdot\nabla v}}&=&-
2\Omega_0{\bld{\hat z}}{\bld{\times v}}+2q\Omega_0x{\bld{ \hat{x}}} -{\bld\nabla} (P/\rho)\ \ ,
\label{eq:bigeom}
\\
\bld{\nabla\cdot v}&=&0  \ ,
\end{eqnarray}
where $\bld{v}$ is the velocity field in the rotating frame and
\be
q\equiv -{d \Omega\over d\ln r}\Big\vert_{r_0}={3\over 2}\Omega_0 \ .
\ee
We use standard Cartesian
 vectorial notation, e.g., $\bld{v}=v_x\bld{\hat{x}}+v_y\bld{\hat{y}}+v_z\bld{\hat{z}}$, etc.
The first term on the right-hand side of equation (\ref{eq:bigeom}) is the Coriolis force, 
and the second is what remains after adding the centrifugal and gravitational forces.

Decomposing  the fluid velocity into
\be
\bld{v}=-qx\bld{\hat{y}}+\bld{u} \ , \label{eq:dec}
\ee
where the first term is the shear flow of the unperturbed disk,  yields
\begin{eqnarray}
\left( {\partial_t}-qx{\partial_y}\right) {\bld u}
+{\bld{u\cdot\nabla u}}
&=&
 2\Omega_0u_y\bld{\hat{x}}-(2\Omega_0-q)u_x\bld{\hat{y}}-\bld{\nabla} (P/\rho) \label{eq:eom2}
\\
{\bld{\nabla\cdot u}} &=&
0 
\label{eq:eom1}
\end{eqnarray}
We shall solve these equations in two dimensions, in the $x-y$ plane.
It is simpler, though,  to work with the curl of equation (\ref{eq:eom2}), which may be expressed in
terms of 
vorticity of $\bld{u}$,
\be
\omega\equiv \bld{\hat{z}\cdot (\nabla\times u})\equiv \partial_x u_y-\partial_yu_x \ ,
\label{eq:om}
\ee
as follows:
\be
(\partial_t-qx\partial_y)\omega
+\bld{u\cdot\nabla }\omega \label{eq:omega} =0 \ , \label{eq:omeq}
\ee
with $\bld{u}$ given by the inverse of equation (\ref{eq:om}):
\be
\bld{u} = \bld{\hat{z}\times\nabla}\nabla^{-2}\omega  \ . \label{eq:uinv}
\ee
Equations (\ref{eq:omeq})-(\ref{eq:uinv}) form a complete set.  
This paper is devoted to solving them.  

Equation (\ref{eq:omeq}) shows that  $\omega$ is 
advected by the
total velocity field $-qx\bld{\hat{y}}+\bld{u}$.  
Vorticity is locally conserved 
in two dimensions (in the absence of dissipation);
$\omega$ is neither 
created nor destroyed, but
is simply shuffled around by the velocity field field that it induces
via equation (\ref{eq:uinv}).
Therefore in 2D investigations one is forced to specify an initial
 vorticity field, and
then to see how it is shuffles itself around. 
In three dimensions, vorticity might be created by turbulence, or
by stirring by planets, or by convection.  But the creation
of vorticity is beyond the scope of this paper.

The vorticity of the unperturbed flow in the rotating frame\footnote{
In the non-rotating frame, the unperturbed disk has total
velocity field
$\bld{v}_{\rm tot}=r\Omega\bld{\hat{\theta}}$, and
the corresponding vorticity is
$\bld{\hat{z}\cdot\nabla\times}\bld{v}_{\rm tot}=d\Omega/d\ln r+2\Omega=
-q+2\Omega_0$ at $r_0$. This differs 
from equation (\ref{eq:vort})
(by $2\Omega_0$) 
because of the change in reference frame.}
is
\be
\bld{\hat{z}\cdot \nabla\times }(-qx\bld{\hat{y}})=-q \ ; \label{eq:vort}
\ee
$q$ sets a scale for the vorticity against which the strength of the perturbation
$\omega$ may be measured.
Throughout this paper, we restrict the perturbed vorticity to be
less than the background vorticity\footnote{This 
restriction precludes us from seeing the shear
(or Kelvin-Helmholtz)
instability discovered by \cite{SSG06}, which requires $|\omega|\gtrsim q$.},
\be
|\omega|\lesssim q \ .
\ee
It is conceivable that hydrodynamic disks are so 
turbulent that they violate this inequality.  But if they reach such 
a state via a nonlinear instability, then the instability would have had to 
act while the inequality held. 
Because of the above inequality, it is tempting to drop the nonlinear term
$\bld{u\cdot\nabla }\omega$ from equation (\ref{eq:omeq}).
However, we shall show that even when $|\omega|\ll q$ the nonlinear term 
is
important provided that $\omega$ is sufficiently elongated in the streamwise direction. 
While we may indeed neglect nonlinear advection in the
$y$-direction ($u_y\partial_y\omega$),
the term $u_x\partial_x\omega$ advects fluid in a direction orthogonal to the background shear. 
The net displacement in $x$ can be significant if $u_x$
acts coherently over a large range in $y$.

\subsection{Equivalence with Non-rotating Linear Shear Flows}

Equations (\ref{eq:omeq})-(\ref{eq:uinv}) also describe the dynamics of 
{\it non-rotating}
incompressible 2D linear shear flows.
To see this, one may discard
the first two terms on the right-hand side of equation (\ref{eq:bigeom}) and make the
decomposition of equation (\ref{eq:dec}), where the unperturbed shear rate
$q$ may now be set to any arbitrary constant.  
Equation (\ref{eq:eom2}) is then reproduced without the 
coriolis terms ($=-2\Omega_0\bld{\hat{z}\times\hat{u}}$).
But the curl of this equation still yields equation (\ref{eq:omeq}), because
in two dimensions the coriolis force
 is the gradient of a scalar,
$-2\bld{\hat{z}\times u}=\bld{\nabla}\nabla^{-2}2\omega$.  Therefore
the coriolis force can be absorbed into a redefinition of the pressure, and
does not affect the dynamics.
 (The sum of centrifugal and gravitational forces $2q\Omega_0\bld{x}$ 
 also does not contribute to the dynamics of $\bld{u}$---it merely
cancels
 the coriolis force due to the background shear $-2\Omega_0\bld{\hat z\times}
 (-qx\bld{\hat{y}})$. But this is true in 3D as well. In fact, this balance sets
 the unperturbed velocity field.)

Rotating and non-rotating incompressible shear flows are equivalent only in two dimensions.
To illustrate how this equivalence breaks down in three dimensions, we examine
 linear axisymmetric waves.
 In non-rotating shear flows (eq. [\ref{eq:eom2}] without the coriolis force), such ``waves''
have
dispersion relation  $\omega=0$, and eigenfunction   $\bld{u}\propto \bld{\hat{y}}$.
 They are static disturbances of the background
flow that merely alter the shear flow's velocity profile.
Rotating shear flows have
dispersion relation
$\omega^2= 2\Omega_0(2\Omega_0-q){k_z^2/ k^2}$.
Therefore two-dimensional modes (with $k_z=0$) have $\omega=0$ whether or not the
flow is rotating.  But axisymmetric modes in rotating flows with $k_z\ne 0$ have a non-zero frequency.
When three dimensional motions are allowed, the coriolis force induces epicyclic-like oscillations.

 \section{Nonlinear Mode Coupling Equation}
\label{sec:fourier}

 We solve equations (\ref{eq:omeq})-(\ref{eq:uinv}) in a box of size $L_x\times L_y$  subject to boundary conditions that are  periodic in $y$, and shifted periodic in $x$ (eq. [\ref{eq:shifted}]).
 Lagrangian coordinates $(\bld{X},T)$
shear with the background flow:
\begin{eqnarray}
Y&\equiv& (y+qtx ){\ \rm mod\ } L_y
\label{eq:y}   \\
(X,T)&\equiv& (x,t) 
\label{eq:xt}
\ \\
\Rightarrow
\partial_T&=&
\partial_t-qx\partial_y \ \ .
\label{eq:dt}
\end{eqnarray} 
The Fourier transform of $\omega$ with respect to Lagrangian coordinates is 
\be
\omega_{J_x,J_y}\equiv 
\int_0^{L_x}
\int_0^{L_y}
{dX\over L_x}
{dY\over L_y}
\omega(\bld{X}) e^{-i\bld{K\cdot X}}  \ , \label{eq:ftdef}
\ee
where 
\be
K_x\equiv {2\pi J_x\over L_x} \ , \ K_y\equiv {2\pi J_y\over L_y} \ , \label{eq:kxky}
\ee
for integer $J_x, J_y$, with inverse
\be
\omega(\bld{X})=\sum_{J_x,J_y}\omega_{J_x,J_y}e^{i\bld{K\cdot X}} \ .
\ee
  Capital letters
($\bld{X}, \bld{K}, J_x, J_y$) denote Lagrangian coordinates, and lower-case letters denote
Eulerian ones.
The transform of equation (\ref{eq:omeq}) gives the nonlinear mode coupling equation,
\beqn
{d{\omega}_{J_x,J_y}\over dT}&=& 
\sum_{J_x',J_y'}
{\omega}_{J_x',J_y'}
{\omega}_{J_x-J_x',J_y-J_y'}
{\bld{\hat{z}\cdot}
\left[\bld{k}({\bld{K-K'}},T)\bld{\times} \bld{k}(\bld{K'},T)\right]\over
|\bld{k}(\bld{K'},T)|^2} 
\nonumber
\\
&=& 
\sum_{J_x',J_y'}
{\omega}_{J_x',J_y'}
{\omega}_{J_x-J_x',J_y-J_y'}
{\bld{\hat{z}\cdot}\left[\bld{K\times K'}\right]
\over|\bld{k}(\bld{K'},T)|^2} 
\ , \label{eq:fourierspace}
\eeqn
where 
 $\bld{K'}$ is related to $(J_x',J_y')$ via
 the analog of equation (\ref{eq:kxky}), and
\be
\bld{k}(\bld{K},T)\equiv\bld{K}+qTK_y{\bld{\hat{x}}}
\label{eq:kxmain}
\ee
is the wavevector with respect to $\bld{x}$. 
Equivalently, since 
a single mode has the spatial dependence 
\be
\exp({i\bld{K\cdot X}})=\exp({i\bld{k}(\bld{K},T)\bld{\cdot x}}) \ ,
\ee
the Fourier transform 
of
$\bld{\nabla}\equiv(\partial_x,\partial_y)$ is obtained
by making the substitution
$\bld{\nabla}\rightarrow i\bld{k}(\bld{K},T)$.
Hence  equation (\ref{eq:fourierspace}) can be read
directly from equation (\ref{eq:omeq}) after replacing 
the Fourier transform 
of $\bld{u}$ with $(-i/k^2)\bld{\hat{z}\times k}$ times the Fourier transform
of $\omega$ (eq. [\ref{eq:uinv}]).

\section{Linear Evolution}

The linearized equation of motion  is $\partial\omega/\partial T=0$, which has the general solution is
\be
\omega=f(X,Y)=
f(x,y+qtx\ {\rm mod} \ L_y )  
\ee
for any function $f$.
A single vorticity wave can be written as
\be
\omega=\bar{\omega}\cos\left(
{\bld{\bar{K}\cdot X}} 
\right) 
={\bar{\omega}}\cos\left(\bld{k}(\bar{\bld{K}},T)\bld{\cdot x}\right)
\ , \label{eq:wave}
\ee
where ${\bar{\omega}}$ and $\bld{\bar{K}}$ are constants.
The velocity fluctuation due to this wave is
\begin{eqnarray}
\bld{u} &=& \bld{\hat{z}\times\nabla}\nabla^{-2}\omega \\
&=&
{\bld{\hat{z}\times k}(\bld{\bar{K}},T)\over |\bld{k}(\bld{\bar{K}},T)|^2}\bar{\omega}\sin
\left(
\bld{\bar{K}\cdot X}
\right) \\
&=&
{(-\bar{K}_y,\bar{K}_x+qT\bar{K}_y)\over (\bar{K}_x+qT\bar{K}_y)^2+\bar{K}_y^2}{\bar{\omega}}\sin
\left(
\bld{\bar{K}\cdot X}
\right) , \label{eq:uswing}
\end{eqnarray}
the amplitude of which
 depends explicitly on $T$.
 
 Equation (\ref{eq:wave}) is not only
a solution of the linearized equation of motion, but it is an exact solution of the nonlinear
equation as well: the nonlinear term $\bld{u\cdot\nabla}\omega$ vanishes for a single wave 
because of incompressibility.

\subsection{Axisymmetric, Swinging, and Radial Waves}

Axisymmetric waves
have $k_y={K}_y=0$: their phasefronts are aligned with 
the $y$-axis, which corresponds to the azimuthal direction in a Keplerian disk. See Figure \ref{fig:kfiglin}.
Axisymmetric waves are independent of time not only in shearing coordinates, but in unsheared ones as well.
``Swinging waves'' have ${K}_y\ne 0$.  At early times ($T\rightarrow-\infty$), they are leading waves: the ratio $k_x/k_y\rightarrow -\infty$, so their phasefronts, though nearly axisymmetric, are tilted slightly  into the first and third quadrants of the $x-y$ plane.
As time evolves, the phasefronts swing first through the radial direction---parallel to the $x$-axis---and then, 
as $T\rightarrow +\infty$ they approach re-alignment with the azimuthal direction as a highly trailing wave.
They are exactly radial at time
\be
T_{\rm rad}(\bld{K})\equiv -{K_x \over qK_y} \ , \label{eq:trad}
\ee
in terms of which
\be
 k_x=K_yq(T-T_{\rm rad}) \ . \label{eq:kxtrad}
\ee
We refer to a swinging wave at time
$T_{\rm rad}$ as a ``radial'' wave, since its
 phasefronts are then aligned with the radial direction.

\includegraphics*[width=9cm]{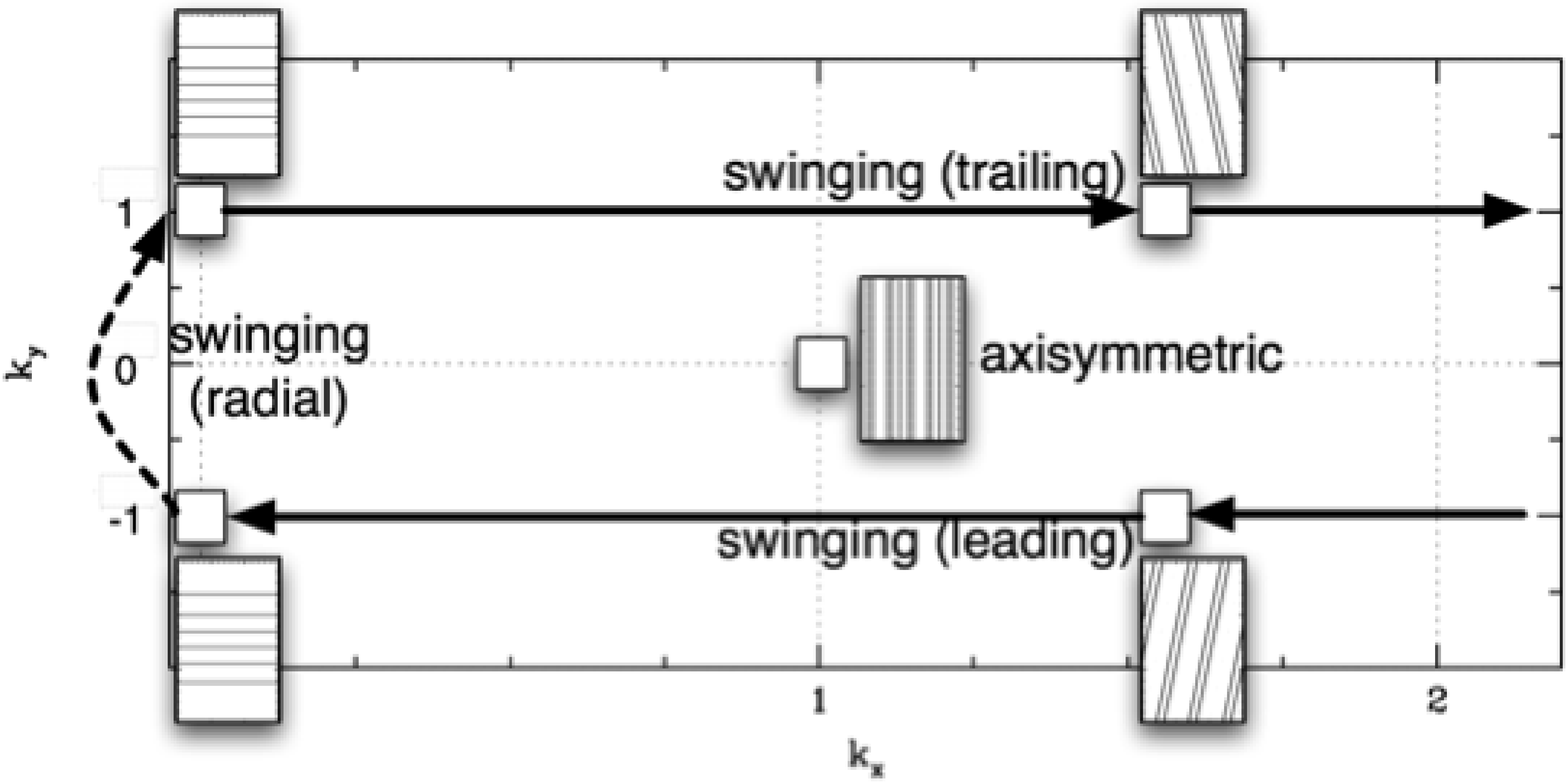}
\figcaption{
{\bf  Linear Evolution in $k$-space:}  
\label{fig:kfiglin}
 Only the half of $k$-space with $k_x>0$ is plotted, the other half being redundant because
a mode with wavevector $-\bld{k}$ is the complex conjugate of a mode with wavevector $\bld{k}$.
Two modes are shown.  The axisymmetric mode is the white square at $(k_x,k_y)=(1,0)$; it does not
move in $k$-space.  The nearby rectangle depicts its contours of constant vorticity in $x-y$ space, and shows
phasefronts aligned with the background shear, parallel to $y$. The second mode is a swinging
mode.  It evolves in time from the lower-right corner to the upper-right one.  Rectangles again
depict phasefronts at the corresponding positions in $k$-space.
 The dashed arrow at $k_x=0$
can be thought of as an instantaneous jump in $k_y$: as a 
 mode with $(k_x,k_y)=(0,-1)$ continues off the plot to negative 
$k_x$, its complex conjugate appears  at $(0,1)$.
}

\subsection{Numerical Simulation: a Single Swinging Wave}

Figure \ref{fig:uxuy}
shows the evolution of $u_x$ and $u_y$ of a swinging wave.
  The points show the output from the
pseudospectral code described in the Appendix, and the curves through the points are given by equation 
(\ref{eq:uswing}).  See the Figure's caption for details.
Figure \ref{fig:contour} shows contours of constant vorticity at three times in this simulation.

\section{Nonlinear Evolution}
\label{sec:ssw}

The nonlinear coupling coefficient between two modes is
time-dependent (eq. [\ref{eq:fourierspace}]), which
at first sight makes the mode coupling problem look forbidding.
But the situation is not quite so dire.
The coupling coefficient
depends on time only through its denominator
\be
|\bld{k}(\bld{K'},T)|^{2}=K_y^{'2}\left(1+q^2(T-T_{\rm rad}({\bld K'}))^2  \right)  \ ,
\ee
which is smallest
at the time when
either of the modes has radial phasefronts ($k_x=0$).
At much earlier or later times
the coupling $\propto T^{-2}$.  Therefore integrated over time, most of the coupling occurs 
around the time that $k_x=0$.  
Because mode couplings occur at essentially discrete
times, the analysis is greatly simplified, as we presently slow.

\includegraphics*[width=9cm]{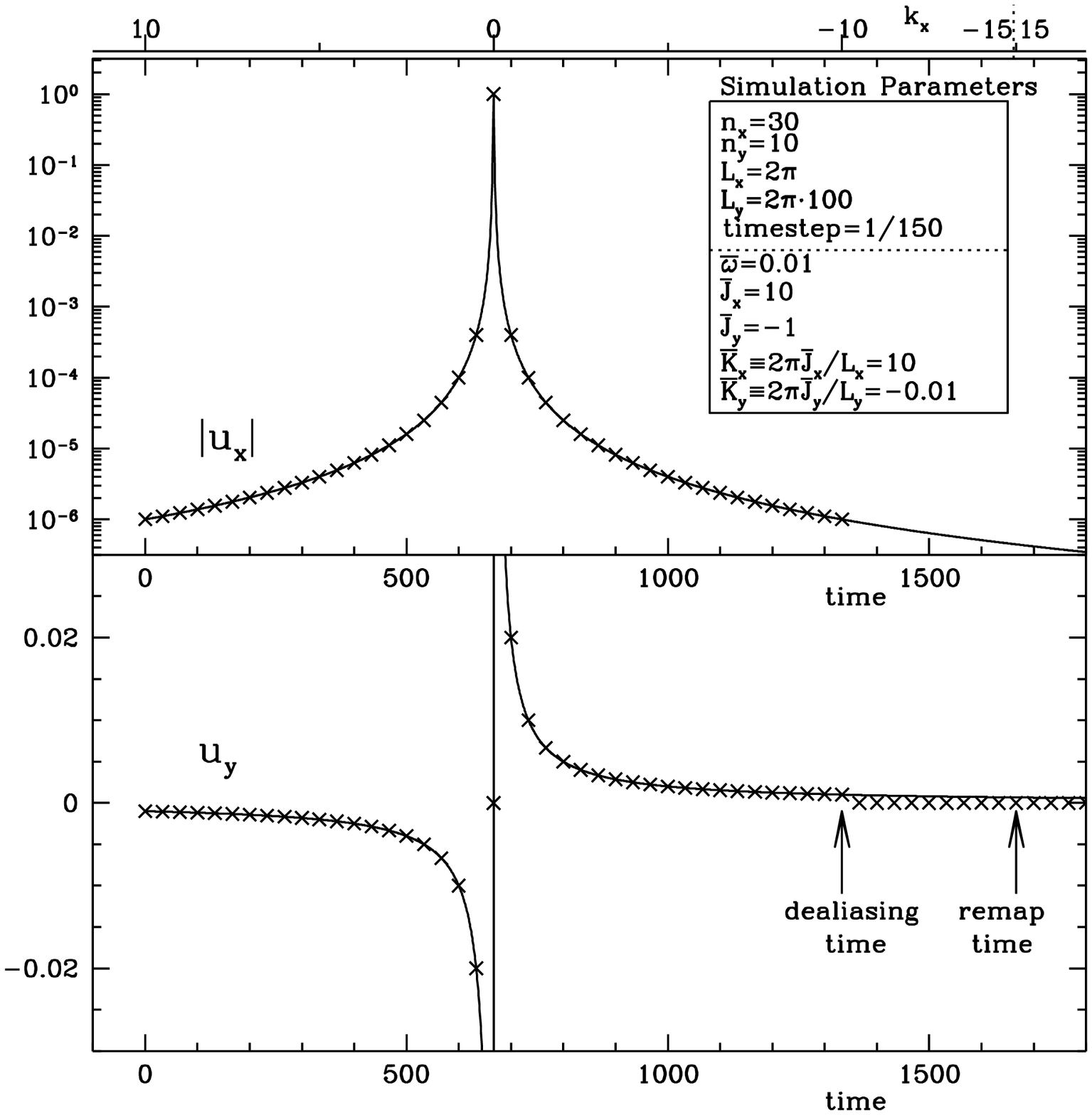}
\figcaption{
{\bf  Evolution of a Single Swinging Wave:}  
\label{fig:uxuy}
At $T=0$, a vorticity wave (eq. [\ref{eq:wave}]) is initialized
with
 parameters as
shown in the figure inset. $\bar{J}_x$ is  the value of $J_x\equiv K_xL_x/2\pi$ 
of the only non-zero mode; and similarly for $\bar{J}_y$; $n_x,n_y$ are the number of gridpoints
in the simulation, and time is measured in units of $\Omega_0^{-1}$.
Points show output from pseudospectral code, properly normalized, i.e., they
show $2{\rm Im}({\bld{{u}}_{10,-1})}$, where 
$\bld{u}_{J_x,J_y}$ is the Fourier transform (as in eq.  [\ref{eq:ftdef}]).
Lines through the points show the amplitude of equation (\ref{eq:uswing}):
  $\bar{\omega}\bar{K}_y/k^2$ in the top panel and $\bar{\omega}k_x/k^2$ in the bottom.
 At time $T_{\rm rad}=-\bar{K}_x/q\bar{K}_y=667$, phasefronts cross through the radial direction.
In the pseudospectral code, dealiasing with the 2/3 rule sets $\bld{{u}}_{10,-1}$ to zero when 
this mode has $|k_x|> n_x/3=10$ (eq.[\ref{eq:dealias}]).
This occurs at time $20/q\vert \bar{K}_y\vert=1333$.  At a later time, when $|k_x|=n_x/2=15$ (which occurs
at time $25/q\vert \bar{K}_y\vert= 1667$) the $k_x$ of this mode is mapped from $k_x=-15$ to $k_x=15$
via the modulus function in equation (\ref{eq:goodkx}).  So this mode once again becomes a leading
mode.  However, since the amplitude of this mode was set to zero by dealiasing, it remains equal to zero. 
}

\subsection{Swinging Waves and a Single Axisymmetric Wave}
\label{sec:asingleaxi}
To begin our investigation of the nonlinear coupling between waves, we consider the evolution 
of small swinging waves in the presence of a much larger axisymmetric wave.
We set 
\be
\omega=2{\omega}_{1,0}\cos(X)+ \omega_{\rm swing}(\bld{X},T) \ , \label{eq:azswing}
\ee
with $|\omega_{\rm swing}|\ll |{\omega}_{1,0}|\ll q$, and $\omega_{1,0}$ a real-valued constant;
the factor of $2$ multiplying $\omega_{1,0}$ has been inserted for consistency with 
equation (\ref{eq:ftdef}).
 We have chosen the length unit so that the axisymmetric mode
has $K_x=1$.

Substituting equation (\ref{eq:azswing}) into equation (\ref{eq:omega}) gives
\be
{\partial\omega_{\rm swing}\over\partial T}
=-2{\omega_{1,0}}\sin(X)
\left(\nabla^{-2} +1
\right)
\partial_y\omega_{\rm swing} \ ,
\label{eq:swing}
\ee
after dropping $\bld{u}_{\rm swing}\bld{\cdot\nabla}\omega_{\rm swing}$.

\includegraphics*[width=9cm]{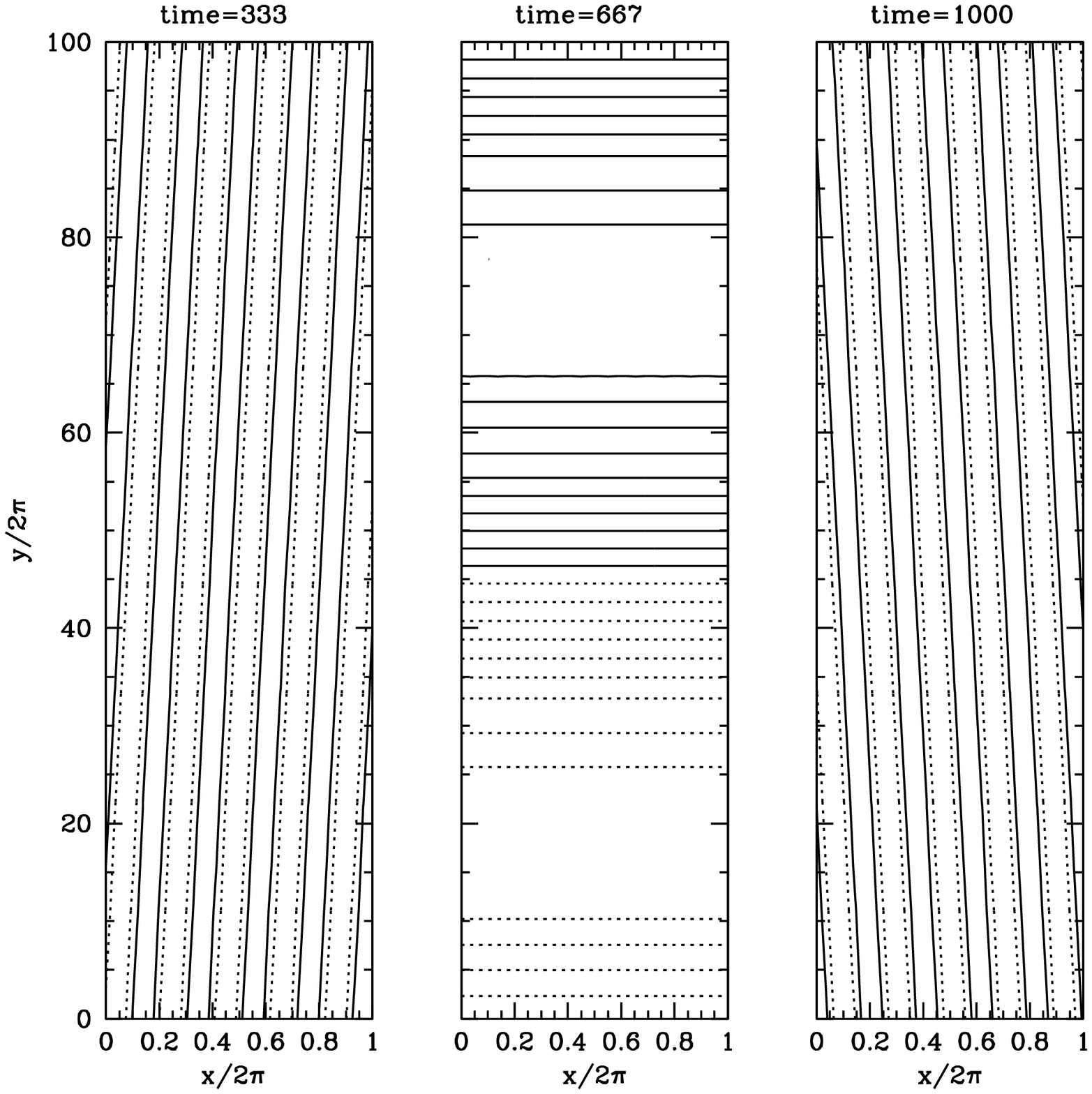}
\figcaption{
{\bf Evolution of a Single Swinging Wave:}  
\label{fig:contour}
Contours of constant vorticity at three times from the simulation described in Figure \ref{fig:uxuy}, showing, respectively, leading, radial, and trailing phasefronts.
}

Equation (\ref{eq:swing}) is linear in $\omega_{\rm swing}$, with a spatially variable coefficient, $\sin(X)$.  A Fourier
mode of $\omega_{\rm swing}$ with wavevector $(K_x,K_y)$ couples with $\sin(X)$ to produce two modes
with wavevectors $(K_x\pm 1, K_y)$.  
Hence a solution of equation (\ref{eq:swing}) is\footnote{
 This is clearly not the complete solution.
But each ``chain'' of modes  represented by equation (\ref{eq:swingchain}), with 
fixed $\bar{K}_y$ and $-\infty<J_x<\infty$,
evolves independently of others.
For the complete solution, one should also 
sum over different ${K}_y$,
and include chains whose $k_x(T=0)$ are not integer multiples of the axisymmetric mode's
$K_x$; 
these chains are also described by equation (\ref{eq:swingchain}), except that their time
origin (when $K_x=k_x$) is offset from $T=0$.  Note also that for simplicity we have not included arbitrary
phases in the arguments of the cosines (i.e., we take mode amplitudes to be real-valued).  It is trivial to include phases; we do so in \S \ref{sec:many}.
 }
\be
\omega_{\rm swing}=
\sum_{J_x=-\infty}^\infty
2\omega_{J_x}(T)\cos
\left(
J_xX-\bar{K}_y Y \label{eq:swingchain}
\right) \ .
\ee
Since we  take $K_y$ of the swinging modes as fixed (with $K_y=-\bar{K}_y<0$), we suppress the corresponding subscript $J_y$ from $\omega_{J_x,J_y}$;  mode amplitudes with a single subscript all label modes with  the same $J_y\ne 0$, where ${K}_y=2\pi J_y/L_y$.
Substituting the above $\omega_{\rm swing}$ into equation (\ref{eq:swing}),
\be
{d\omega_{J_x}\over dT}=-{\bar{K}_y {\omega}_{1,0}}
\left(
{\omega_{J_x-1}
\over k_{J_x-1}^2}
-
{\omega_{J_x+1}
\over k_{J_x+1}^2}
\right) \ , 
\label{eq:3mode}
\ee
where
\begin{eqnarray}
k_{J_x}^2&\equiv& (J_x-qT\bar{K}_y)^2+\bar{K}_y^2 \\ 
&=&
\bar{K}_y^2(1+q^2(T-T_{{\rm rad},J_x})^2) \ ,
\end{eqnarray}
and 
\be
T_{{\rm rad},J_x}\equiv T_{\rm rad}(J_x,-\bar{K}_y)={J_x\over q\bar{K}_y} \label{eq:tradjx}
\ee
(eq. [\ref{eq:trad}]).  In writing equation (\ref{eq:3mode}), we kept only the $\nabla^{-2}$ term from
the right-hand side of equation (\ref{eq:swing}).  
To include the full $\nabla^{-2}+1$, one should replace the
$k_{J_x\pm 1}^{-2}$ in equation (\ref{eq:3mode})
with $k_{J_x\pm 1}^{-2}-1$;
we show below that the omitted terms are small (eq. [\ref{eq:tildechi}]).
The $\nabla^{-2}$ term is due to the advection of the axisymmetric
wave by the swinging ones, i.e., it is 
$- \bld{u}_{\rm swing}{\bld{\cdot\nabla}}2\omega_{1,0}\cos\left(X\right)$.
The  dropped term
represents
advection of the swinging waves by the axisymmetric one.

If $\bar{K}_y\ll 1$, then the squared wavevectors that appear in 
the denominator in  equation (\ref{eq:3mode}) reach
very small values when the corresponding mode is radial.  At this time, the mode
strongly influences its two neighboring modes.
We shall see that we can often approximate the interaction between different swinging waves
 as occurring solely when one of the waves passes through the radial direction.

As an example, at $T=0$ we initialize $\omega_{\rm swing}$ with a single leading wave:
\be
\omega_{\rm swing}(\bld{X},T=0)= 2\epsilon \cos(X-\bar{K}_y Y) \ , \label{eq:ic}   \
\ee
i.e., with $\omega_1=\epsilon$, and $\omega_{J_x}=0$ for $J_x\ne 1$.
We choose
\begin{eqnarray}
0<\bar{K}_y&\ll &1  \\
  \epsilon&\ll& |\omega_{1,0}| \ ,
\end{eqnarray}
and follow the evolution governed by equation (\ref{eq:3mode}).
See Figure \ref{fig:kfig}.

\includegraphics*[width=9cm]{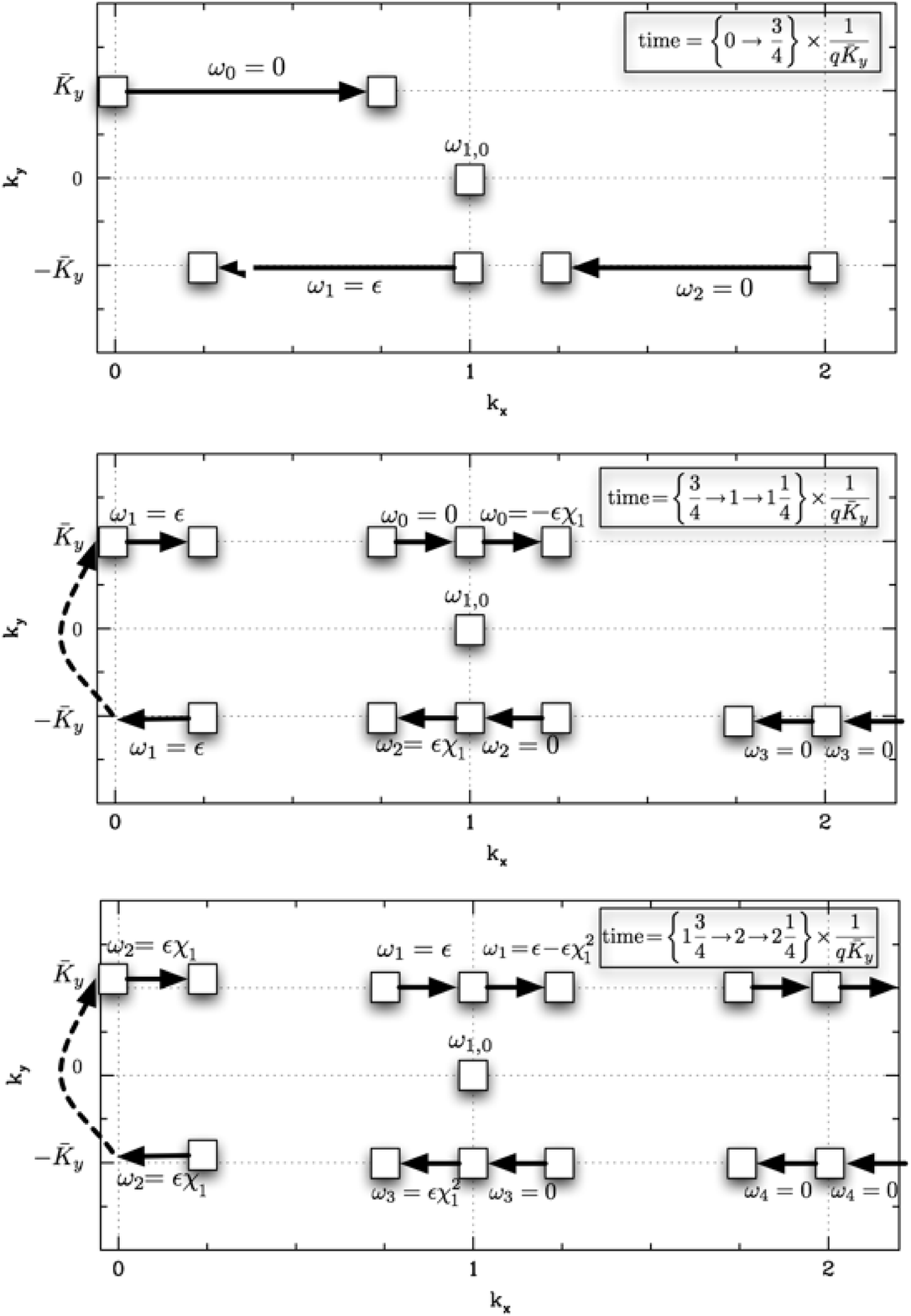}
\figcaption{
{\bf  Early-Time Evolution in $k$-space:
\label{fig:kfig}
The top panel shows four modes between the times $0$ and $3/4q\bar{K}_y$.  For example, the square at $(k_x,k_y)=(1,-\bar{K}_y)$ represents $\omega_1$ at time $T=0$; the 
connecting arrow
shows this mode's trajectory in $k$-space until time $3/4q\bar{K}_y$.
The middle panel continues the evolution through time  $T_{\rm rad,1}=1/q\bar{K}_y$,
when $\omega_1$ is radial, and $\omega_0$ and $\omega_2$  interact strongly
 with the radial and axisymmetric modes. At this time, $\omega_2$ changes from
$0$ to $\epsilon\chi_1$, and
 $\omega_0$ changes from $0$ to $-\epsilon\chi_1$.  The bottom panel follows the evolution
through time $T_{\rm rad,2}=2/q\bar{K}_y$.
}}

At first, the only non-zero swinging mode is $\omega_1$. This mode influences only two others:
$\omega_2$ and $\omega_0$.  So at early times, $\omega_2$ approximately satisfies
\begin{eqnarray}
{d\omega_2\over dT}&=&-{\bar{K}_y\omega_{1,0}}{\omega_1\over  k_{1}^2} \label{eq:om2dot} \\ 
&=&
-{\omega_{1,0}\over \bar{K}_y}
{\omega_1\over 
1+q^2(T-T_{{\rm rad},1})^2
}
\end{eqnarray}
Integrating this from time $T=0$ through  $T_{rad,1}=1/(q\bar{K}_y)$, we see that most of the integral comes from within a
time $\sim 1/q$ of $T_{rad,1}$, when the phasefronts of $\omega_1$ are within
around $\pm45^o$ to the radial direction.  So we can extend the integration to the domain
$-\infty<T<\infty$ to obtain shortly after time 
 $T_{rad,1}$, i.e. at time $\sim(1+q\bar{K}_y)T_{rad,1}$,
\be
\omega_{2}
=\omega_{1}\chi_1 =\epsilon \chi_1 \  \label{eq:do2}
\ee
where
\be
\chi_1\equiv -{\pi\over q\bar{K}_y}{\omega_{1,0}} \  ,  \label{eq:chi1}
\ee
 is the dimensionless parameter that controls the linear evolution of $\omega_{\rm swing}$.
The behavior of $\omega_0$ at time $T_{rad,1}$ is similar to $\omega_2$, with
$\omega_{0} =-\epsilon\chi_1$ shortly after $T_{rad,1}$.

 In the top panel of Figure \ref{fig:kfig}, the modes initially evolve linearly,
with constant amplitude and $k_x=K_x-qTK_y$.
 Before time $T_{\rm rad,1}$, both
$\omega_1$ and $\omega_2$ are leading modes (their $k_x/k_y<0$), and $\omega_0$ is trailing.
But when $\omega_1$ becomes radial and its $k_x=0$, the amplitudes of modes $\omega_0$ and $\omega_2$ change abruptly (Figure \ref{fig:kfig}, middle panel).   Qualitatively, the velocity field of the $\omega_1$
mode when it becomes radial is $\bld{u}^{\rm rad}=-\bld{\hat{x}}(2\epsilon/\bar{K}_y)\sin(\bar{K}_yy)$ (eq. [\ref{eq:uswing}]).
Since it
takes a time $\sim 1/q$  to swing through the radial direction, the corresponding displacement
field is $\bld{\xi}^{\rm rad} 
\sim \bld{u}^{\rm rad}/q$, which advects the axisymmetric mode, changing it by an amount
$-\bld{\xi}^{\rm rad}\bld{\cdot\nabla}2\omega_{1,0}\cos(X)\sim -2\epsilon(\omega_{1,0}/q\bar{K}_y)\left(
\cos(x-\bar{K}_yy)-\cos(x+\bar{K}_yy)\right)$.  Hence  the $\omega_2$ mode, which has the spatial dependence
$\cos\left(x-\bar{K}_yy\right)$ at this time, changes its amplitude by
 $\sim -\epsilon(\omega_{1,0}/q\bar{K}_y)$, as in equation (\ref{eq:do2}); similarly, $\omega_0$  is changed  by an equal and opposite amount.

After time $T_{\rm rad,1}$, the next time of interest is $T_{\rm rad,2}=2/q\bar{K}_y$, when the  
$\omega_2$ mode is radial with  $\omega_2=\epsilon\chi_1$
(Figure \ref{fig:kfig}, bottom panel).
The evolution of $\omega_3$ around this time is given by the analog of equation
(\ref{eq:om2dot}), i.e., $d\omega_3/dT=-\bar{K}_y\omega_{1,0}\omega_2/k_2^2$.
Integrating through time $T_{\rm rad,2}$ yields $\omega_3=\omega_2\chi_1=\epsilon\chi_1^2$ shortly
after $T_{\rm rad,2}$.  Similarly, $\omega_1$ changes by an equal and opposite amount.

Extrapolating to later times, it is clear that the amplitude of the mode that crosses through the radial direction
is 
\be
\omega_{J_x}=\epsilon \chi_1^{J_x-1} 
{\ \rm at\ time\ }T_{{\rm rad},J_x}={J_x\over q\bar{K}_y}
    \ . \label{eq:chi}
\ee
As time progresses, the amplitudes of the radial waves
  grow exponentially if $|\chi_1|>1$; conversely, they
   decay exponentially if $|\chi_1|<1$.  
At marginal stability,
\be
\kappa\equiv
\bar{K}_y\Big\vert_{\rm marginal\ stability}={\pi |\omega_{1,0}|\over q} \ . \label{eq:kappa1}
\ee
 (If we do not set $K_x=1$ for the axisymmetric mode then 
 $\kappa=  \pi |K_{x,\rm axi} \omega_{1,0}/q|$).
Since $|\chi_1|=\kappa/\bky$, equation (\ref{eq:chi}) shows
\be
|\omega_{J_x}|\propto \exp\left({|\omega_{1,0}|}T_{{\rm rad},J_x}\pi (\bky/\kappa)\ln(\kappa/\bky)\right) \ ,
\label{eq:grsin}
\ee
 and hence the growth rate is very small both when $\bky\rightarrow \kappa$ and when
$\bky\rightarrow 0$; it is fastest for $\bky=\kappa/e=0.368\kappa$.

\subsection{Numerical Simulation}

Figures
 \ref{fig:2modes}  and \ref{fig:2modespp} show results from two pseudospectral simulations.
 At $T=0$, the vorticity field was given by equations (\ref{eq:azswing}) and (\ref{eq:ic}), i.e., 
 $\omega(T=0)=2\omega_{1,0}\cos(X)+2\epsilon\cos\left(X-\bar{K}_yY\right)$.
One of the simulations had $\chi_1=-1.1$, and the other had $\chi_1=-0.9$.
 In Figure \ref{fig:2modes}, we plot
 the amplitudes of the first eight modes that pass through the radial direction in the
 $\chi_1=-1.1$ simulation.
 Most of the time the amplitudes remain constant.
Only  at times $T_{{\rm rad},J_x}$ do the amplitudes of modes $J_x\pm 1$ change abruptly; see also Figure \ref{fig:kfig}.

\includegraphics*[width=9cm]{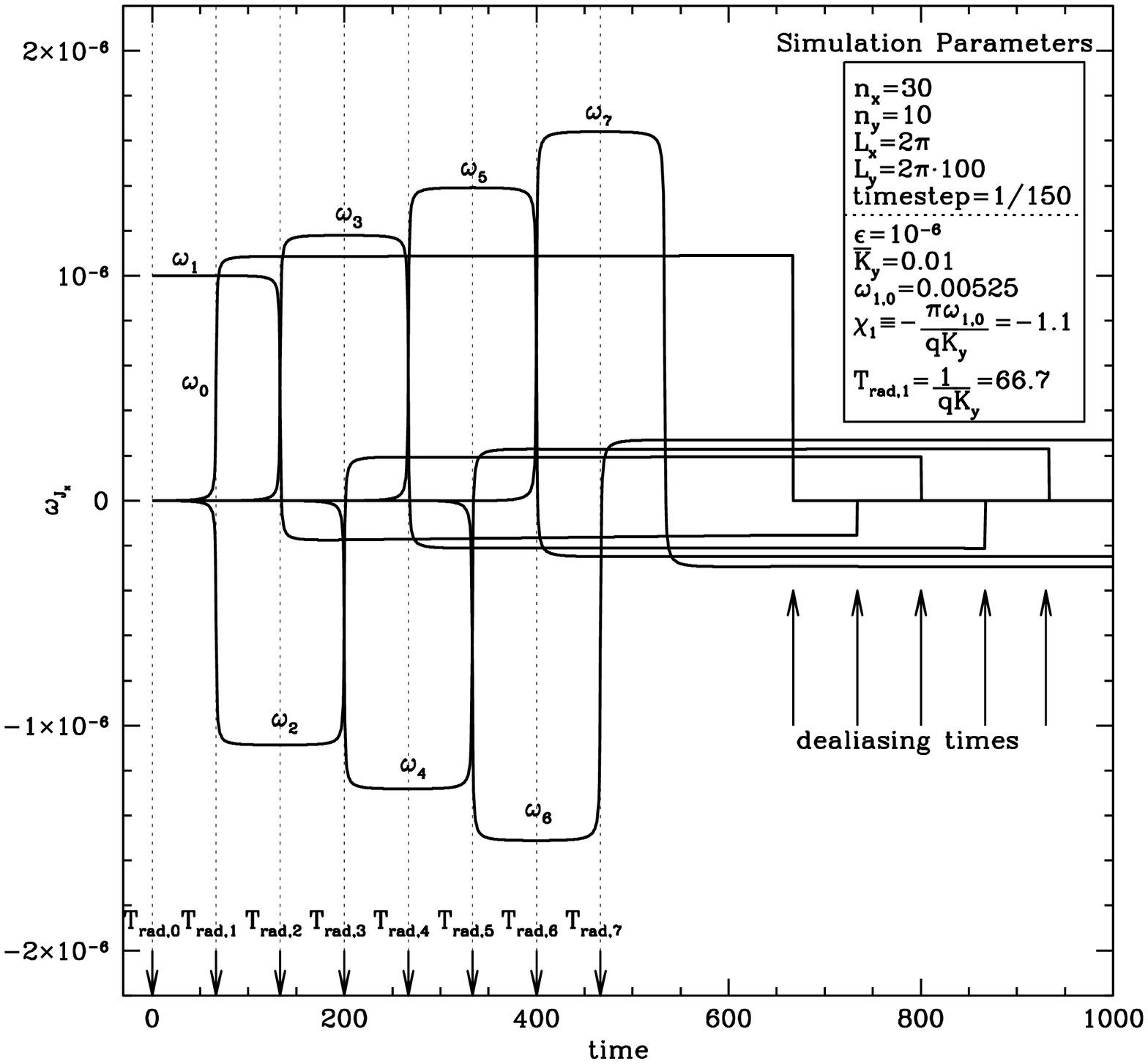}
\figcaption{
{\bf  Swinging Waves in the Presence of a Large Axisymmetric Wave:
\label{fig:2modes}
Pseudospectral simulation of the fully nonlinear equations of motion.  
The vorticity field was initially
 $\omega(T=0)=2\omega_{1,0}\cos(X)+2\epsilon\cos\left(X-\bar{K}_yY\right)$
(eqs. [\ref{eq:azswing}] and [\ref{eq:ic}]);
 parameters are listed in inset.
Curves show $\omega_{J_x}={\rm Re}({\omega}_{J_x,-1})$. Dealiasing is described in \S \ref{sec:numerical}.}}

 The points in Figure \ref{fig:2modespp} show the amplitudes of radial modes at each time
 that a mode is precisely radial.
The line through the points is the theoretical prediction.
(Of course, only at the discrete times $T_{{\rm rad},J_x}$, for integer $J_x$, is there a radial mode; the theoretical prediction is plotted as a line for clarity.)  The theoretical prediction is given by equation (\ref{eq:chi}), but
with a slightly modified $\chi_1$.
  In particular, in deriving
equation (\ref{eq:chi}), we made two approximations that are 
applicable as $\bar{K}_y\rightarrow 0$; we shall now include the $O(\bar{K}_y)$ corrections.
First, we integrated equation (\ref{eq:om2dot}) from $-\infty$ to $\infty$, whereas we should
have integrated from $\sim 0$ to $\sim 2/q\bar{K}_y$; second, instead of $\omega_1/k_1^2$ in equation (\ref{eq:om2dot}), we should have $\omega_1(1/k_1^2-1)$.  Both corrections can be absorbed into a redefinition
of the dimensionless parameter,
\be
\tilde{\chi}_1\equiv\chi_1\cdot (1-{4\over\pi}\bar{K}_y) \ ; \label{eq:tildechi}
\ee
 $\tilde{\chi}_1$ should be inserted in equation (\ref{eq:chi}) in place of $\chi_1$.
As long as $\bar{K}_y\ll 1$, this correction is small.
Nonetheless, accumulated corrections can be substantial if $\tilde{\chi}_1$ is 
 exponentiated many times.  For example, in the top panel of Figure {\ref{fig:2modespp}}, using
$\chi_1=-1.1$ instead of $\tilde{\chi}_1=-1.086$ would overpredict the amplitude of the last plotted point
by a factor of 2.

\includegraphics*[width=9cm]{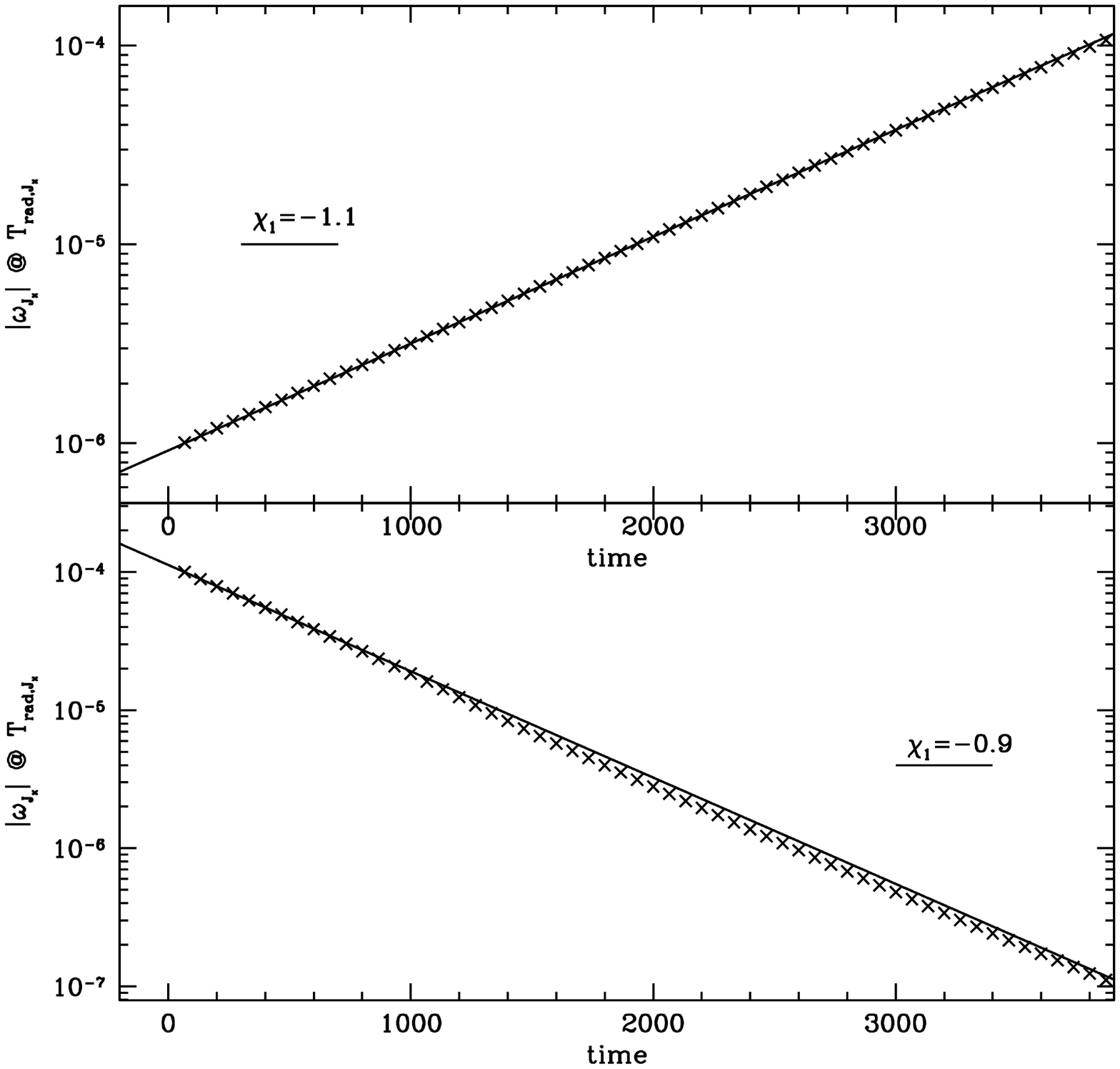}
\figcaption{
{\bf  Amplitudes of Radial Modes:
\label{fig:2modespp}
Points in the top panel show results from the pseudospectral simulation described in Figure \ref{fig:2modes}, extended to later times.  Each point shows the amplitude of the mode that is radial at that time.
The line through the points shows the theoretical prediction
 $\epsilon\vert\tilde{\chi_1}\vert^{q\bar{K}_yT-1}$, as in eq. [\ref{eq:chi}], but with
$\tilde{\chi}_1=-1.086$  in place of $\chi_1=-1.1$;  see eq. [\ref{eq:tildechi}].
Bottom panel shows results from a simulation identical to top panel's, except with
 $\omega_{1,0} = 0.0043\Rightarrow\chi_1=-0.9$ and $\epsilon=10^{-4}$.  Note that
 the last point in both panels has $J_x=58$, even though the simulations only have 
 $n_x/2=15$ modes in the $x$-direction: modes are ``recycled''  by the remapping
 that is performed by the modulus operator (eq. [\ref{eq:goodkx}]).}
} 

\subsection{Quasi-Eulerian Notation}

Analysis of  the coupling between Fourier modes is simpler when a quasi-Eulerian 
notation is used; we introduce this notation here.
Instead of $\omega_{J_x,J_y}$, where the subscripts  label the mode's
Lagrangian wavevector $\bld{K}$, we label a mode's amplitude 
\be
\omega^{j_x,j_y}
\ee
when the mode's
Eulerian wavevector $\bld{k}(\bld{K},T)$ satisfies
\be
k_x={2\pi\over L_x}j_x {\ \ \rm and}\ \ k_y={2\pi\over L_y}j_y \ , \label{eq:smallj}
\ee
where $j_x$ and $j_y$ are any integers. 
Since $k_x$ is time dependent, equation (\ref{eq:smallj}) only applies at discrete times.  But we shall also label a mode's amplitude with $\omega^{j_x,j_y}$ at times shortly before, and shortly after, its $\bld{k}$ satisfies equation (\ref{eq:smallj});  see Figure {\ref{fig:euler}}.  (To be more precise, one might say that  a mode whose $k_x$ increases in time has its label switched from $\omega^{j_x,j_y}$ to $\omega^{j_x+1,j_y}$ when its $k_x$ passes through
$(2\pi/L_x)(j_x+1/2)$;  but this increased precision is unnecessary.)

\includegraphics*[width=9cm]{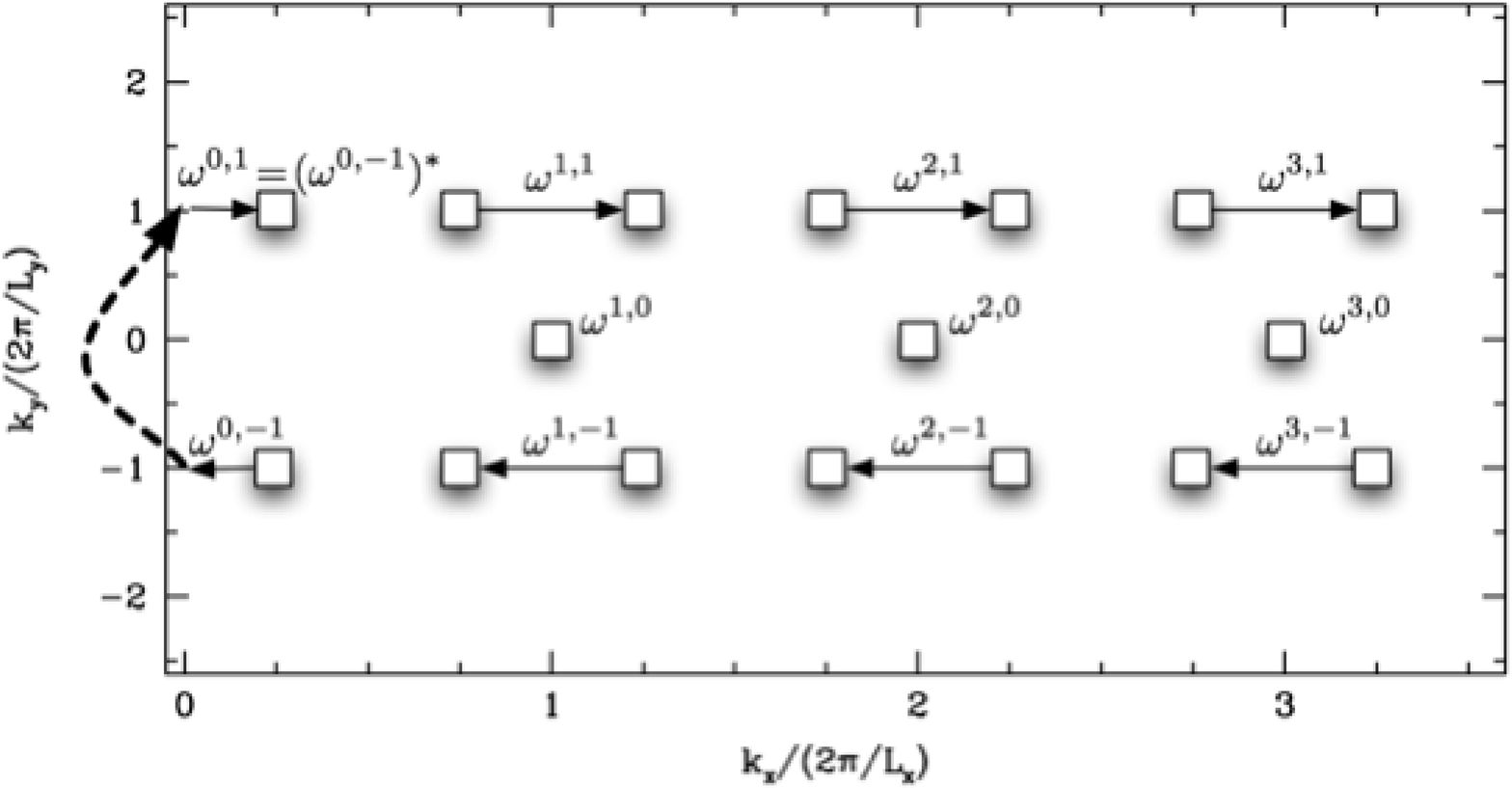}
\figcaption{
{\bf  Quasi-Eulerian Notation:
\label{fig:euler}
Squares represent the wavevectors $\bld{k}$ of a number of modes at two successive times, $T=T_{\rm rad}-1/4q\bar{K}_y$ and $T=T_{\rm rad}+1/4q\bar{K}_y$, where $T_{\rm rad}$ is the radial time of the mode labelled
$\omega^{0,-1}$ in the Figure.  
Modes are labelled by superscripts $j_x,j_y$ representing the nearest
$k_x,k_y$.
Axisymmetric modes
have $k_y=0$. Because their $\bld{k}$ does not evolve, 
their quasi-Eulerian indices are the same as their Lagrangian ones.
}}

\subsection{Many Axisymmetric Waves}
\label{sec:many}

We have shown above that a single axisymmetric wave is unstable to interactions with
swinging waves whose $\bar{K}_y$ are sufficiently small that $|\chi_1|>1$.  In the present
subsection, we show how to generalize this result to consider many axisymmetric waves, and so re-derive
standard results regarding the instability of shear flows  with inflection points in the limit that
$|\omega|\ll q$ and that the vorticity is elongated in the streamwise direction.

We generalize equation (\ref{eq:azswing}) to
\be
\omega=\omega_{\rm axi}(x)+\omega_{\rm swing}(\bld{X},T) \ , \label{eq:ommany}
\ee
where $\omega_{\rm axi}$ is an arbitrary axisymmetric function,
\be
\omega_{\rm axi}(x)=\sum_{j_x=-\infty}^{\infty}\omega^{j_x,0}\exp\left(i j_x x\right) \ ,
\label{eq:omegaaxi}
\ee
with $\omega^{j_x,0}$ arbitrary constants and $L_x=2\pi$.
 We seek solutions to linear order in 
$\omega_{\rm swing}$.
No new values of $k_y$ can be generated by mode couplings because swinging modes must couple with axisymmetric modes, which have $k_y=0$.
We take $\omega_{\rm swing}$ to have $k_y=\pm \bar{K}_y$, where
 $\bar{K}_y>0$, and $\bky$ is much smaller than the typical $k_x$ of $\omega_{\rm axi}$.
 Equation (\ref{eq:fourierspace}) gives the equations of motion in Lagrangian notation.  
 We switch to quasi-Eulerian notation by replacing subscripts with superscripts, and capitalized indices with lower-case ones.  As  in \S \ref{sec:asingleaxi}, for small $\bar{K}_y$ most of the coupling between modes occurs when 
one of the modes is radial, i.e., when the term $|\bld{k}(\bld{K'},T)|^2$ in the denominator of equation 
(\ref{eq:fourierspace}) nearly vanishes.  So we take $\omega_{J_x',J_y'}$ in this equation to be the nearly radial mode
 $\omega^{0,\pm 1}$
 (setting $L_y=2\pi/\bky$ so that $j_y=\pm 1$),
 and, from  equations (\ref{eq:fourierspace})  and (\ref{eq:kxtrad}),
\be
{d\omega^{j_x,-1}\over dT}=
-\omega^{0,-1}\omega^{j_x,0}{j_x\over \bar{K}_y}
{1\over 1+q^2(T-T_{\rm{rad}})^2} \ , \label{eq:domdt}
\ee
where $T_{\rm{rad}}$ is the time when 
$\omega^{0,-1}$ is radial.
Equation (\ref{eq:domdt})  generalizes the first term in equation
(\ref{eq:3mode}) to arbitrary $j_x$ and complex mode amplitudes.
We integrate  from $T=-\infty$ to $+\infty$ because most of the integral comes from within a time $\sim 1/q$ of $T_{\rm{rad}}$.\footnote{
We are assuming that the amplitude of a swinging
mode hardly changes as it  crosses through the radial direction, i.e., for $-\bky\lesssim k_x\lesssim \bky$. 
This assumption 
follows from the requirement that $\bky$ be much smaller than the typical $k_x$ of $\omega_{\rm axi}$.}
The 
change in $\omega^{j_x,-1}$ at this time is
\be
\Delta\omega^{j_x,-1}
=\chi_{j_x}\omega^{0,-1}
\ ,
\ee
where
\be
\chi_{j_x}\equiv -{\pi\over q \bky}j_x\omega^{j_x,0} \ .
\ee
The mode that is labelled $\omega^{0,-1}$ at time $T_{\rm{rad}}$ was, at the earlier
 time $T_{\rm{rad}}-1/q\bky$,  labelled $\omega^{1,-1}$.  At that earlier time, its change due to the then-radial mode was $\chi_1\omega^{0,-1}\big\vert_{T_{\rm{rad}}-1/q\bky}$.  Extrapolating to still earlier times, we conclude
\be
\omega^{0,-1}\Big\vert_{T=J/q\bky}=\sum_{j_x'=1}^\infty \chi_{j_x'}\omega^{0,-1}
\Big\vert_{T=(J-j_x')/q\bky} \ , \label{eq:omrad}
\ee
where $J$ is an integer.
Solutions of this difference equation have the form
\be
\omega^{0,-1}\Big\vert_{T=J/q\bky}= {\rm constant}\times z^{J} \ , \label{eq:diffsol}
\ee
with $z$ determined by the roots of the characteristic equation
\be
1=\sum_{j_x'=1}^\infty \chi_{j_x'}z^{-j_x'} \ . \label{eq:dr}
\ee
This is the dispersion relation; unstable modes have $|z|>1$.
The eigenfunction is, generalizing equation (\ref{eq:omrad}),\footnote{
To be more precise, the left-hand side of equation (\ref{eq:fourierefn}) should be evaluated not  precisely at
time $J/q\bky$, but slightly sooner---before this swinging mode has been altered
by the radial mode. But this distinction is unimportant when 
the typical 
$j_x$ of $\omega_{\rm axi}$ is much larger than unity, in which case
$j_x$ may be treated as a continuous
variable.}
\beqn
\omega^{j_x,-1}\Big\vert_{T=J/q\bky}
&=&
\sum_{j_x'=j_x+1}^\infty
\chi_{j_x'}\omega^{0,-1}\Big\vert_{T=(J+j_x-j_x')/q\bky}
\label{eq:fourierefn}
\\
&=&
{\rm constant} \times z^{J}
\sum_{j_x'=j_x+1}^\infty
\chi_{j_x'}z^{j_x-j_x'} \ . \label{eq:efn}
\eeqn

At marginal stability,
\be
z=e^{-i\beta} \ , \label{eq:zms}
\ee
for a real-valued $\beta$, and the dispersion relation (eq. [\ref{eq:dr}]) is
\be
1=
{1\over 2q\bky}
\int_{-\infty}^\infty 
dx
{d\omega_{\rm axi}/dx\over x-\beta}
+
i {\pi\over 2 q \bky}
{d\omega_{\rm axi}\over dx}
\Big\vert_{x=\beta}
  \ , \label{eq:condition}
\ee
after changing the sum over $j_x'$ to an integral, and extending the box-size $L_x$ to $\infty$.
The  imaginary part of the dispersion relation shows that $\beta$ must be chosen at a zero of 
$d\omega_{\rm axi}/dx$; the real part gives the marginally stable $\bky$:
\beqn
 \kappa&\equiv&
\bky\Big\vert_{\rm marginal\ stability} \\
&=&{1\over 2 q}
\int_{-\infty}^{\infty}dx
{
d\omega_{\rm axi}/dx
\over
x-\beta
}
\Big\vert_{\beta={\rm \ zero\ of\ }d\omega_{\rm axi}/dx}
\label{eq:kymarginal}
\eeqn
In \S \ref{sec:asingleaxi}, we had $\omega_{\rm axi}=2\omega_{1,0}\cos x$; inserting this into the above integral gives $\kappa=\pi|\omega_{1,0}|/q$, as in
 equation (\ref{eq:kappa1}).
For an arbitrary profile of
$\omega_{\rm axi}$,
if the right-hand side of equation (\ref{eq:kymarginal}) is negative for each $\beta$ that is a zero 
of $d\omega_{\rm axi}/dx$, then that profile is stable to all small perturbations.  Conversely, if the right-hand side is positive for any $\beta$, then there are unstable modes with $\bky<\kappa$. This is the general stability criterion. 
Equation (\ref{eq:kymarginal}) is derived by \cite{Gill65}, although using an entirely different
method.
That shear flows can become unstable only when the velocity
field has an inflection point
(i.e., when $d\omega_{\rm axi}/dx=0$) is known 
as Rayleigh's inflection point theorem \citep{DR04}.

For an unstable mode,
\be
z=e^{\alpha-i\beta} \ , \ \ \alpha>0 \ ,
\ee
and the dispersion relation is
\be
1=
{1\over 2q\bky}
\int_{-\infty}^\infty 
dx
{d\omega_{\rm axi}/dx\over x-\beta-i\alpha} \ , \label{eq:dr2}
\ee
which reduces to equation (\ref{eq:condition}) as $\alpha\rightarrow 0$.

Since the eigenfunction is a convolution in Fourier space (eq. [\ref{eq:efn}]), it is simply expressed 
in real-space:
\be
\omega_{\rm swing}=
e^{\alpha q T\bky}
 {d\omega_{\rm axi}\over dx}
\cdot
{\rm Re}
\left\{
{\rm const}
{
e^{-i\bky(y+\beta qT\bky )}
\over
1+i(x-\beta)/\alpha
}
\right\} \ , \label{eq:efnreal}
\ee
where the constant is an arbitrary complex number, and we have set $J=qT\bky$.
A more transparent way of writing the eigenfunction is to combine it with $\omega_{\rm axi}$
\be
\omega=\omega_{\rm axi}+\omega_{\rm swing} = \omega_{\rm axi}(x+\xi(x,y,T)) \ ,
\ee
where the displacement field $\xi$ is given by
\be
\xi(x,y,T)=\xi(0,y_0,0)\times 
e^{\alpha q T \bky}
{\rm Re}
\left\{
{1\over 1+ix/\alpha}
e^{-i\bky (y-y_0)}
\right\} \ ; \label{eq:xi}
\ee
$y_0$ is an arbitrary constant, and 
we have  set $\beta=0$, which 
can be done by changing the origin of the $x$-axis.

The exponential growth rate is $\gamma=\alpha q \bky$. Note that $\alpha$ has the physical interpretation that fluid at 
$x=\pm \alpha$ is advected by the background flow $-qx\bld{\hat{y}}$ a distance $1/\bky$ in the growth time
$1/\gamma$.   Fluid at $|x|\gg \alpha$ is advected many wavelengths in a growth time; this phase mixing implies that the eigenfunction is cut off at $|x|\gg \alpha$.\footnote{
Although $\omega_{\rm swing}$ is cut off at $|x|\gtrsim \alpha$, $u_{x,\rm swing}$ extends further, to 
$|x|\sim \pm 1/\bky$; see \S \ref{sec:piece}.}

\subsection{Numerical Simulation}

We use the pseudospectral code to simulate the equations of motion with $\omega_{\rm axi}$ given by a Gaussian profile,
\be
\omega_{\rm axi}(x)=-\bar{\omega}\exp\left(-x^2/s^2\right) \ , \label{eq:omaxsim}
\ee
with $s\bky\ll 1$ and $\bar{\omega}>0$.
This profile is somewhat unrealistic because it does not integrate to zero.  Hence it
produces a velocity field $u_y$ that is infinite
in extent.  Nonetheless, we show in \S \ref{sec:sunk} that profiles that do not integrate
to zero behave similarly to those that do.

The marginally stable wavenumber is (eq. [\ref{eq:kymarginal}])
\be
\kappa=\sqrt{\pi}{\bar{\omega}\over qs} \ . \label{eq:gausskappa}
\ee
The imaginary and real parts of the dispersion relation (eq. [\ref{eq:dr2}])\
imply
\beqn
\beta&=&0
\\
1&=&
{\bar{\omega}\over  q\bky s^2}
\int_{-\infty}^\infty 
dx
{x^2
\over
x^2+\alpha^2
}e^{-x^2/s^2} \\
&=&
{\kappa\over \bky}
\left(
1-
\sqrt{\pi}
{\alpha\over s}
e^{(\alpha/s)^2}
{\rm erfc}(\alpha/s)
\right)
 \label{eq:invert}
\eeqn
Solving the latter equation for $\alpha$ gives the exponential growth rate 
\be
\gamma=\alpha q\bky=\bar{\omega}f({\bky\over \kappa})
 \label{eq:growthrate}
\ee
 where  $f(\bky/\kappa)$
is plotted
in the top panel of Figure \ref{fig:growthrate}.  The three points in that panel show results from numerical simulations;  see caption for details.  
The fastest growing mode has $\bky/\kappa=0.299$
and growth rate $=0.435\bar{\omega}$.  
The bottom panel shows that
 when our assumption $s\bky \ll 1$ is no longer valid, the growth rate differs from equation (\ref{eq:growthrate}).

In Fourier-space, the eigenfunction at a fixed time is 
\beqn
\omega^{j_x,-1}
&=&
{\rm constant}
\times
e^{-(s j_x/2)^2}
\times
\nonumber
\\
&&\left(
1-
\sqrt{\pi}
{\alpha\over s}
e^{(\alpha/s+sj_x/2)^2}
{\rm erfc}
\left(
\alpha/s+sj_x/2
\right)
\right)
  \label{fig:efngauss}
\eeqn
after changing the sum in equation (\ref{eq:efn}) to an integral. In Figure \ref{fig:efn}, we show that
this agrees with the output from a numerical simulation.

Figure \ref{fig:contours} shows contours of constant $\omega$ at two times, and shows
that theory agrees with simulation.   At later times than those shown, 
the vorticity tends to wrap up into a vortex, with large spatial variations in $\omega$.
To simulate this with the pseudospectral code, we must introduce an explicit viscosity,
which smooths out the variations in $\omega$.   But the highly nonlinear state can be simulated
with no viscosity with a Lagrangian code; we shall do so in \S \ref{sec:twonum}.

\includegraphics*[width=9cm]{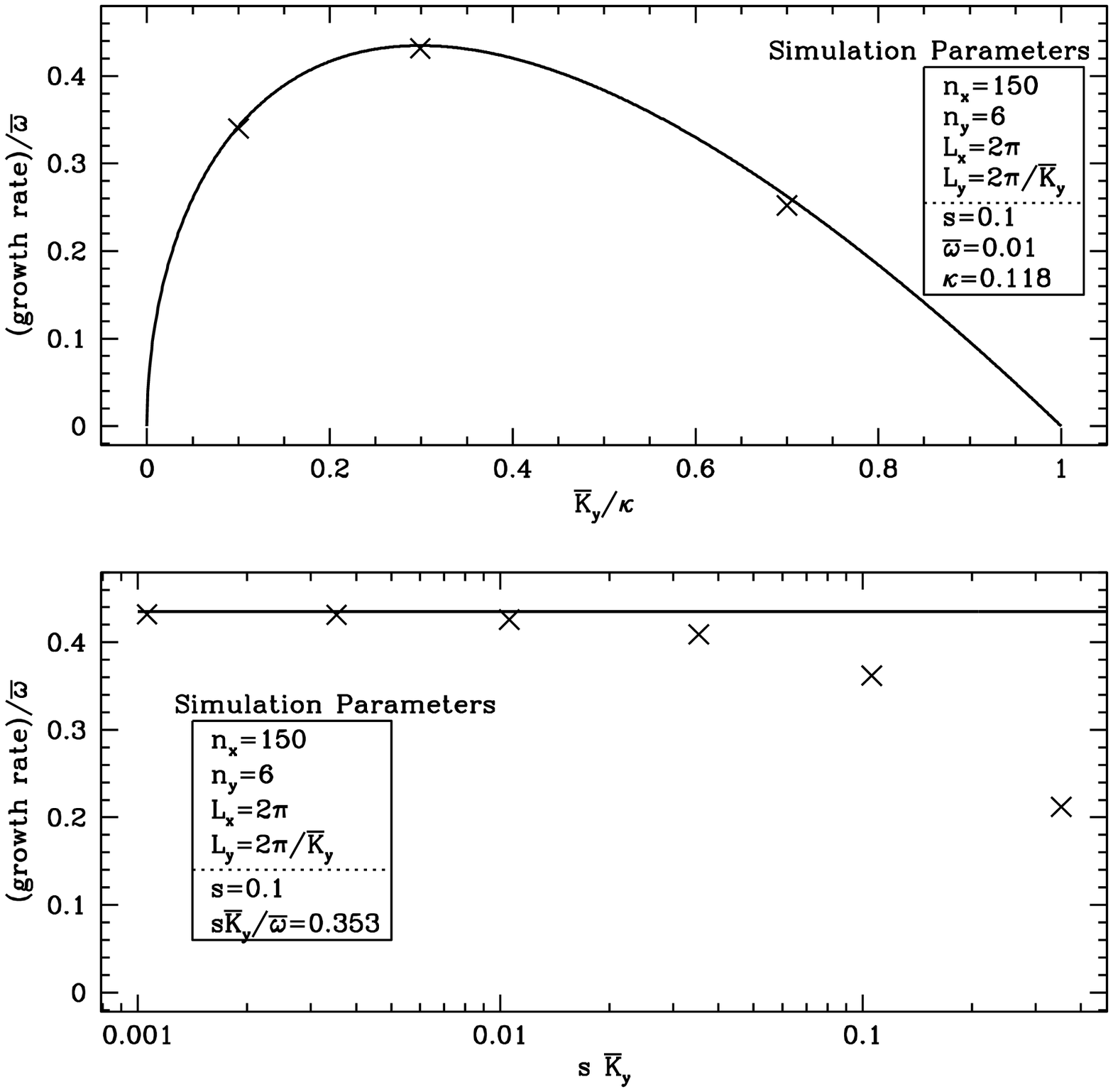}
\figcaption{
{\bf Exponential Growth Rate:
\label{fig:growthrate}
Curve in top panel shows growth rate obtained by solving equation (\ref{eq:invert}) for $\alpha/s$ with Mathematica.
Points show results from three simulations with varying $\bky$, but otherwise identical.  In the simulations, $\omega_{\rm axi}$ was
initialized as $-\bar{\omega}\left(\exp(-x^2/s^2)-\sqrt{\pi}s/L_x\right)$, which differs from  equation (\ref{eq:omaxsim}) by the addition of a constant term that eliminates the zero-frequency component.
  $\omega_{\rm swing}$ was initialized as $\epsilon\cos(\bky y)$, with $\epsilon=10^{-6}$.  The growth rate was extracted from each simulation by plotting the amplitudes of the radial modes versus time (as in Figure \ref{fig:2modespp}), and fitting with an exponential.   For the fastest growing mode,
  $\bky/\kappa=0.299$.
  In the bottom panel, we 
  show results from six simulations with fixed
  $\bky/\kappa=0.299\Rightarrow
   s\bky/\bar{\omega}=0.353$, and fixed $s$, but varying $\bky$ and $\bar{\omega}$.   Equation (\ref{eq:growthrate}) predicts that  the
  growth rate should be equal to 0.435 $\bar{\omega}$,
   independent of $s\bky$.  Numerical results (points in figure) agree with this prediction  for
   $s\bar{K}_y\lesssim 0.01$.
The theory leading to equation (\ref{eq:growthrate}) is only applicable for $s\bky\ll 1$.  
}}

\includegraphics*[width=9cm]{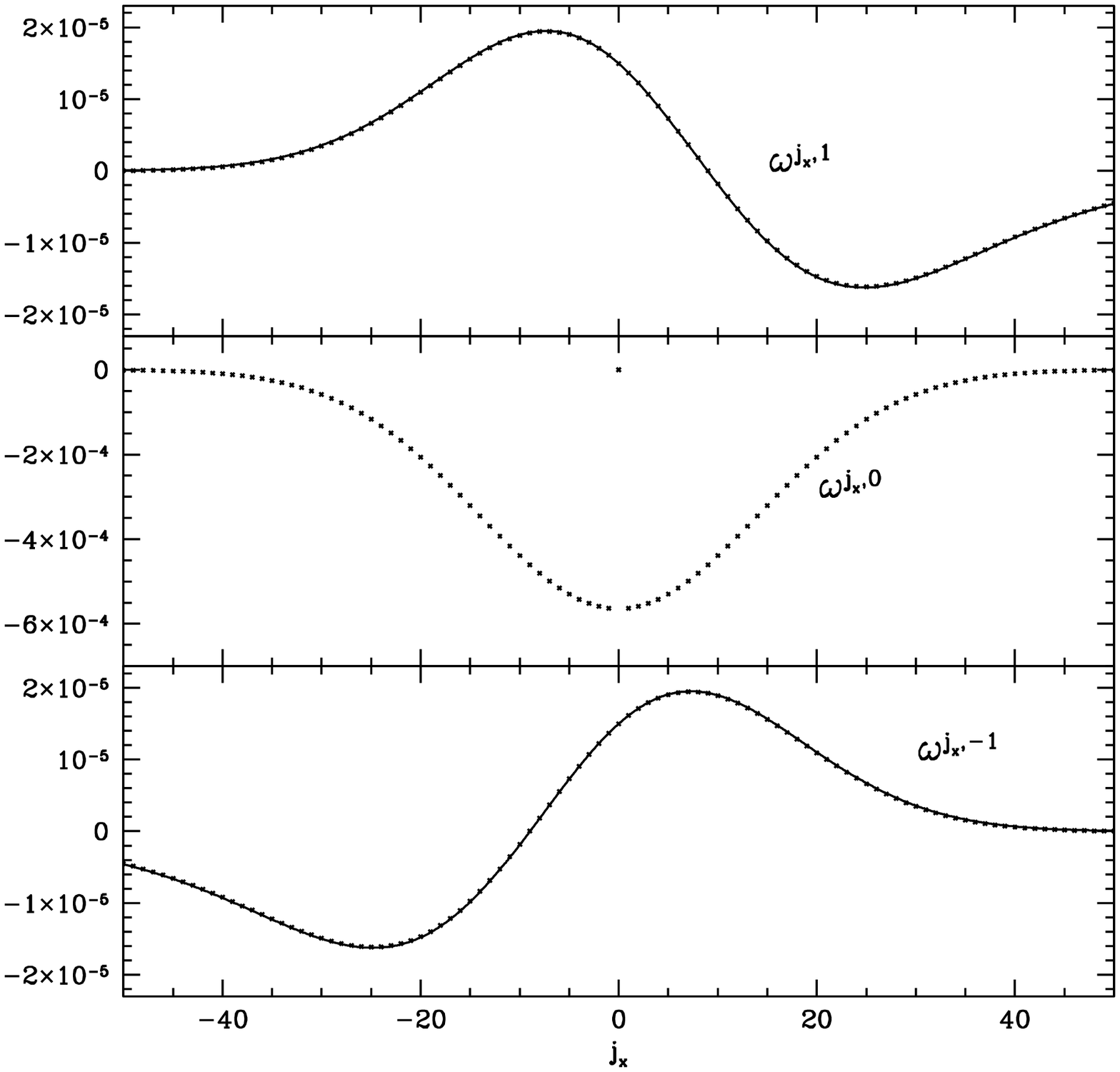}
\figcaption{
{\bf  Vorticity in Fourier-Space:
\label{fig:efn}
Points show output from the simulation described in Figure \ref{fig:growthrate} that had  $\bar{\omega}=0.01$, $s=0.1$, $\bky/\kappa=0.299$, and hence $\alpha=0.082$.
The vorticity shown was output at time $T=1208$.
Top and bottom panels depict the eigenfunction.  By the reality condition $\omega^{j_x,1}=(\omega^{-j_x,-1})^*$.
The curve through the points in these two panels is equation (\ref{fig:efngauss}), with the normalization chosen
to fit the points. The agreement between theory and simulation is quite good.
The middle panel simply reflects the initial $\omega_{\rm axi}$, i.e., it is the Fourier transform of 
equation (\ref{eq:omaxsim}) (with no $j_x=0$ component, as described in Figure \ref{fig:growthrate}).
}}
\includegraphics*[width=9cm]{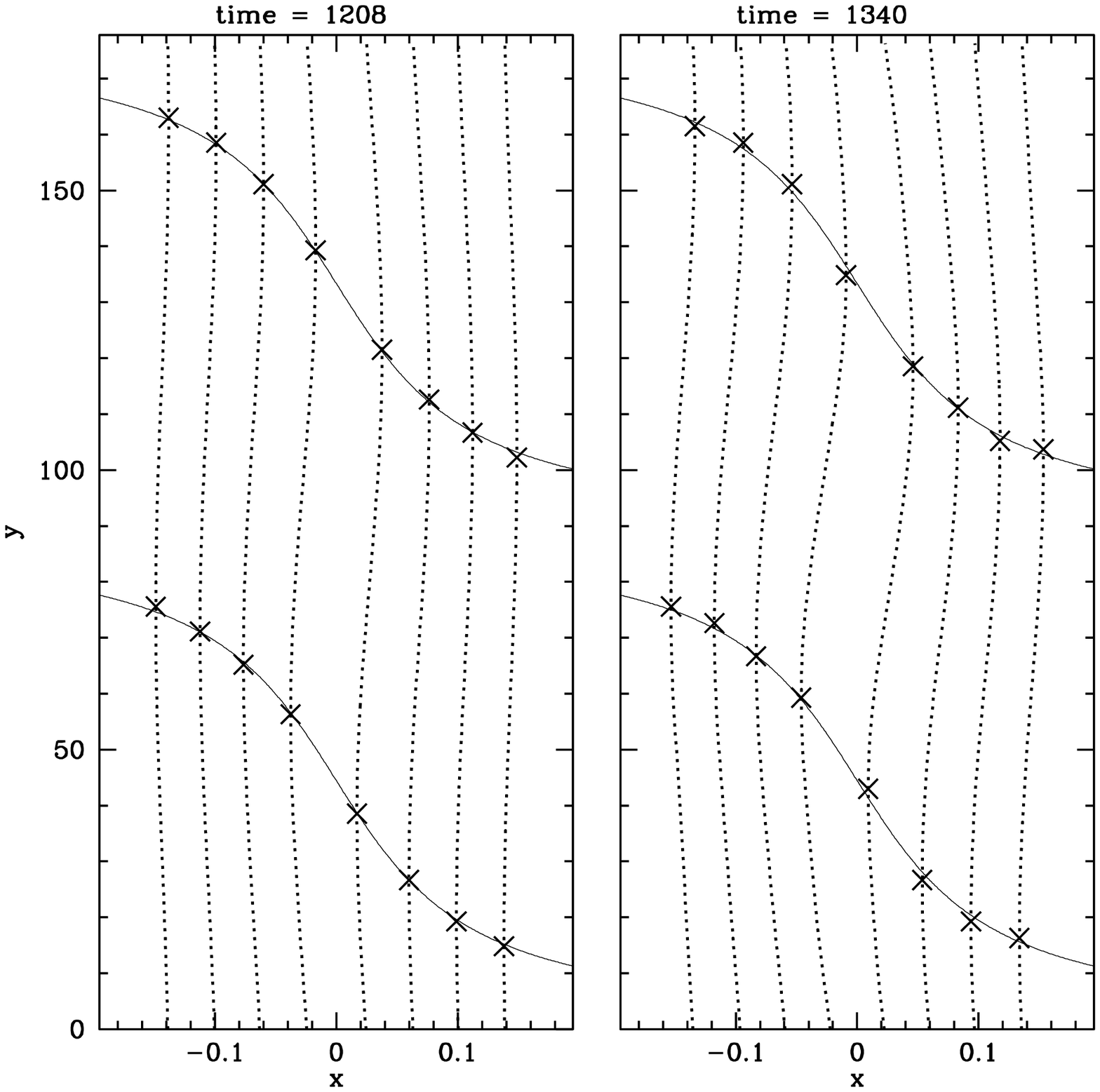}
\figcaption{
{\bf Contour Plots at Two Times:
\label{fig:contours}
Dotted lines trace contours of constant $\omega=\omega_{\rm axi}+\omega_{\rm swing}$ from a
simulation  identical to the one described in Figure \ref{fig:efn}, but with $n_x=600$ instead of $150$.
 To compare  with the theoretical prediction (eq. [\ref{eq:efnreal}], or equivalently,
eq. [\ref{eq:xi}]), we plot two points on each contour 
where the contour's $x$-position is extremal.  From equation (\ref{eq:xi}), 
 the extrema should be at
$y_{\rm ext}=(\phi-\tan^{-1}(x/\alpha))/\bky$, where $\phi=\pi/2$ and $3\pi/2$.  We plot  these two $y_{\rm ext}$ as solid lines, with $\alpha=0.082$ (as obtained from Figure \ref{fig:growthrate}).
 The lines nearly go through the points, implying good agreement.
}}

\section{Shear Instability in Real Space}
\label{sec:piece}

We seek a better understanding of the instability discussed 
in \S \ref{sec:ssw}.
The instability of shear flows with inflection points in the streamwise velocity field---or equivalently,
with $d\omega_{\rm axi}/dx=0$ (eq. [\ref{eq:kymarginal}])---is well known \citep{DR04}. However, we have not
found in the literature
a simple yet quantitatively accurate explanation for why it occurs.
Why are only certain shear flows unstable, and why only for sufficiently 
small
wavenumbers?
We give a dynamical explanation for a top-hat profile in \S \ref{sec:tophat}, and 
 an explanation  based on momentum and energy considerations in 
  \S \ref{sec:momen}.  
  We also simulate the development
of the instability into the highly nonlinear regime in \S \ref{sec:twonum}.
We generalize these results from a top-hat profile to an arbitrary one in \S \ref{subsec:infinite}, 
and  present another simulation in \S \ref{sec:sunk}.

\subsection{Top-Hat Vorticity Profile}
\label{sec:tophat}

We consider in some detail the  dynamics when the unperturbed $\omega$ is given
by the top-hat profile
\be
 \omega_{\rm unp}
  =
 \left\{
 \begin{array}
 {r@{\quad,\quad}l}
 \mu & |x|<s \\
 0 & |x|>s 
 \end{array}  
 \right. 
 \ ,
 \label{eq:tophat}
\ee
where $\mu$ and $s$  are constants, with $s>0$.
The total unperturbed vorticity is $-q+\omega_{\rm unp}$, and we assume
that $|\mu|\ll q$.  The top-hat profile is perhaps the simplest profile that exhibits instability.
It is not a realistic profile, both because 
real fluids are not discontinuous, and
because it produces  a $u_y$-field that is infinite in extent.
But 
 once the results for the top-hat are in hand, it trivial to extend them
to profiles that
do not suffer from these defects.
We do so in \S\S \ref{subsec:infinite}-\ref{sec:sunk}, where we show that more realistic profiles
behave similarly to the top-hat.

\includegraphics*[width=9cm]{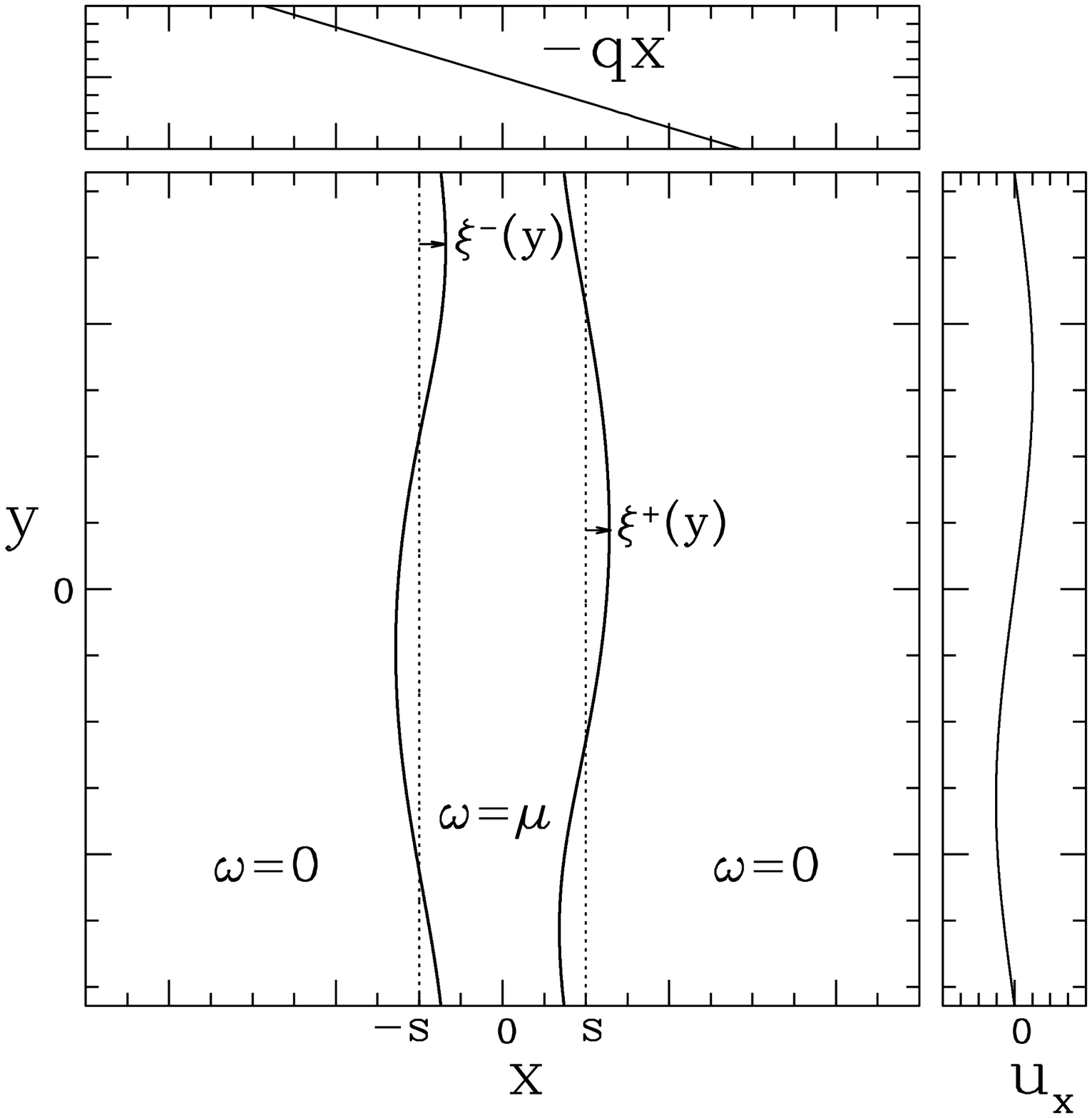}
\figcaption{
{\bf Perturbed Top-Hat Profile:
Main panel shows contours of constant vorticity.  Dotted lines are for the unperturbed
profile (eq. [\ref{eq:tophat}]), and solid lines are perturbed.
Top and right panels show the velocity fields that advect the contours.
The 
top panel shows the amplitude of the background shear profile $-qx{\bld{\hat{y}}}$, and the
right one shows $u_x$ as a function of $y$.  As long as inequality (\ref{eq:smallky})
is satisfied, $u_x$ can be taken to be independent of $x$.  
$u_y$ is not depicted because 
 advection in the $y$ direction is dominated by the background shear.
The data for this figure come from the simulation described in \S \ref{sec:twonum}
at time=4000. This simulation has $\mu<0$ and $|k_y|<|\mu|/qs$, so
that instability results.
Spatial and velocity  axes are not to scale.
\label{fig:tophat}
}}

 We perturb the top-hat  by displacing the two contours at $x_-$ and $x_+$ by 
$\xi^-(y,t)$ and $\xi^+(y,t)$, as in  Figure \ref{fig:tophat}, and seek equations of motion
for the contours
\citep[``contour dynamics,'' e.g., ][]{Pullin92}. 
We can adopt this partially
Lagrangian description  as long as a contour does not fold back upon itself;
otherwise,
$\xi$ becomes multiple-valued, and a fully Lagrangian or an Eulerian description
would be preferable.
 Although tracking contours is simpler than
the mode coupling method of earlier sections, mode coupling can readily be generalized to three
dimensions, whereas contour dynamics cannot: by tracking the contours, we rely on the fact that $\omega$ is locally conserved,
 which is only true in two dimensions.
We still assume that the dynamics occur in a box that
is periodic in $y$ with size $L_y$; but we now assume $L_x$ is sufficiently
large that the boundary conditions in $x$ are unimportant. (From eq.
[\ref{eq:uxky1}] below, it suffices to take $L_x\gg 1/k_y$ for all $k_y$.)
We shall make use of Fourier transforms, but only in the $y$-dimension, with
\begin{eqnarray}
&\omega_{k_y}(x)\equiv &\int_{-L_y/2}^{L_y/2}\omega(x,y)e^{-ik_yy}{dy\over L_y} \\ 
&{\Rightarrow}\omega(x,y)=&\sum_{k_y}\omega_{k_y}(x)e^{ik_yy} 
\end{eqnarray}
and similarly for $u_{x,k_y}(x)$, $\xi_{k_y}$ etc., where $k_y=(2\pi/L_y)\times$ integer.

We shall see that instability can only occur when 
$|k_y|s< |\mu|/q\ll 1$; therefore we assume at the outset
that perturbations are elongated in the streamwise direction,
\be
|k_y|s\ll 1 \ . \label{eq:smallky}
\ee

The contours are advected by the total velocity field
 $-qx{\bld{\hat{y}}}+\bld{u}$ (eq. [\ref{eq:omeq}]), implying
\be
(\partial_t \mp qs\partial_y)\xi^\pm = u_x \ , 
\ee 
after dropping the nonlinear terms $(-q\xi^\pm+u_y)\partial_y\xi^\pm$.
In Fourier-space,
\be
\left({d\over dt}\mp qs i k_y\right)\xi_{k_y}^\pm=u_{x,k_y} \label{eq:xikyeq}
\ee
We need not distinguish the $u_x$ that advects $\xi^+$ from the
one that advects $\xi^-$, as we presently show.

To find $u_{x,k_y}$, we have 
 $u_x=-\partial_y\nabla^{-2}\omega$ (eq. [\ref{eq:uinv}]), or 
  \beqn
 u_{x,k_y}&=&{i\over 2}{\rm sgn}(k_y)\int \omega_{k_y}(x')e^{-|k_y(x-x')|}dx' \label{eq:uxky1} \\
 &\simeq& {i\over 2}\mu(\xi^+_{k_y}-\xi^-_{k_y}){\rm sgn}(k_y)
 \  , {\rm for}\  |x|\lesssim s \label{eq:xi1}
 \eeqn
  To derive the latter relation,
  the vorticity near the ``step'' at $x=s$ may be expanded as
 $
 \omega={\rm const}-\mu H(x-s-\xi^+) 
 \simeq \mu\xi^+(y,t)\delta(x-s) +f(x)
 $
 where $H(x)$ is Heaviside's step function, $\delta(x)=dH/dx$ is Dirac's delta function,
  and $f(x)$ denotes $y$-independent
 terms that
 may be discarded
 because ${\rm sgn}(0)=0$. The  step  at $x=-s$ contributes with the opposite sign.
The exponential in equation (\ref{eq:uxky1}) may be dropped because of inequality (\ref{eq:smallky}).

Even though $\omega$ varies rapidly with $x$, the $u_x$ that it induces is independent
of $x$. 
(More precisely, $u_x$ decays exponentially in $x$ on lengthscale $1/|k_y|\gg s$.)
This result, which will simplify the analysis, 
corresponds to our earlier finding that swinging waves only contribute
significantly when their phasefronts are radial.  
More generally, one observes from equation (\ref{eq:uxky1}) 
that any vorticity distribution that is elongated in the streamwise direction induces
a velocity field that is much smoother in the $x$-direction than is $\omega$.
The convolution smooths out variations in $x$ with a smoothing length 
of $1/|k_y|$;
hence $u_x$ varies in $x$---as well as in $y$---on the scale $1/|k_y|$.
One can further argue on dimensional grounds that, in general, elongated vorticity distributions
 produce a velocity field with amplitude
$u_x\sim \omega\xi$.  When $\xi=0$, one must have $u_x=0$ by symmetry
considerations. And although there is a dimensionless combination
$k_ys$ (where
$s$ would represent the lengthscale of $\omega$ in the $x$-direction),
$u_x$ cannot depend on $s$ because it smooths out $\omega$ on scales much larger
than $s$.

The simple equations 
 (\ref{eq:xikyeq}) and (\ref{eq:xi1}) 
embody all the dynamics of the initial stages of the shear instability.
To solve
them,
we insert the trial solution $\xi^\pm\propto e^{\gamma t}$,
which produces for the  eigenvector and eigenvalue
\beqn
\xi^\pm_{k_y}&\propto&{1\over \gamma \mp iqsk_y} 
\label{eq:efngood}
\\
\gamma^2&=& -(qsk_y)^2-qs|k_y|\mu
\label{eq:gammagood}
\eeqn
The behavior of the growth rate $\gamma$ is very similar to that found for 
previously considered vorticity profiles
 (eqs. [\ref{eq:grsin}], [\ref{eq:growthrate}] and Fig. \ref{fig:growthrate}).
When $\mu>0$, i.e. when the perturbed vorticity 
and the background vorticity $-q$ have opposite signs, the flow is stable
for all $k_y$.
When $\mu<0$, the flow is unstable provided that
\be
|k_y|<\kappa\equiv
{|\mu|\over qs} \ .
\label{eq:kapkap}
\ee
The maximum growth rate is $\gamma_{\rm max}=|{\mu}|/2$, which
occurs at $|k_y|=\kappa/2$, and $\gamma$ falls to zero as either
$k_y\rightarrow 0$ or $k_y\rightarrow\kappa$.
 
We may now begin to appreciate how the shear instability works in real space.
Depending on the phases of the sine waves $\xi^+(y)$ and $\xi^-(y)$, the velocity field $u_x$ that 
they induce can tend to amplify them.  
Figure \ref{fig:tophat}  shows a growing mode in an unstable shear flow that has 
$\mu<0$, so that the vorticity in the strip down the middle has the same sign as
the background vorticity $-q$.
The strip bulges
 in its central region,  where  $|y|\lesssim  L_y/4$, and because 
 of the sign of $\omega$, the bulge
 tends to produce a swirling velocity around itself with the same
sense as the background shear.  Therefore it produces a positive
$u_x$ in the top of the figure, and a negative one at the bottom, 
as depicted in the right panel.
Furthermore, the strip 
is tilted into the top-right and bottom-left quadrants of the $x-y$ plane.
Although the tilt might not be obvious at first glance, if one compares
the vorticity along any line of fixed $|y|$ with that at $-|y|$, one observes
that the former is to the right of the latter.
Because the induced $u_x$ 
pulls the top half of the strip further to the right and the bottom half further to the left,
it tends to amplify the tilt. 
In sum, $u_x$ converts a bulge into a tilt.   When $\mu<0$, the 
tilt is into the shear, having the same sense as a leading wave.

To complete the picture, consider the effect of the
background shear,
which carries
$\xi^+$ downwards at speed $qs$, and $\xi^-$ upwards at the same speed.
As long as the vorticity strip is tilted into the shear, 
the background shear converts a tilt into a bulge, thus completing the loop and
allowing for exponential growth.

Based on the above picture, we
may estimate the fastest growth time as $\gamma_{\rm max}^{-1}\sim \xi^+/u_x\sim 1/|\mu|$
 (eq. [\ref{eq:xi1}]).   Advection of the
two contours
by the background shear can dephase them before growth can occur. 
Since  the dephasing time is
$t_{\rm dephase}\sim 1/|qsk_y|$, it will be shorter than the fastest growth time
if  $|k_y|$ is sufficiently large.
In other words, if $|k_y|\gtrsim  |\mu|/qs$, the flow is stable,
as found above.

\subsection{Momentum and Energy for the Top-Hat}
\label{sec:momen}
 
 It is
instructive to see how  momentum and energy interact to trigger the shear instability.
Not only does this give a deeper understanding of why the instability operates, it
also gives a sense of the nonlinear outcome of the instability, since it quantifies
the driving force behind it.  Furthermore, the interaction between 
 momentum and energy is  generally
important in accretion disks.  Understanding how this works in a nontrivial incompressible
flow might be helpful when considering more complicated effects, such as three dimensional
motions, 
baroclinicity,  or magnetohydrodynamics.

We  first derive expressions for the momentum and energy in terms of the dynamical variables
$\xi^\pm$, and show that the equations of motion
 (\ref{eq:xikyeq}) and (\ref{eq:xi1}) 
exactly conserve these quantities.  
Perturbations to the top-hat profile have
 $y$-momentum (per unit $L_y$ and setting $\rho=1$) 
\beqn
{M}  &\equiv& \int (u_y-u_y\vert_{\xi^\pm=0})  dx{dy/ L_y}
\\
&=&\int  (\overline{u}_y-\overline{u}_y\vert_{\xi^\pm=0}) dx  \ ,
 \label{eq:mom}
\eeqn
where
the bar  denotes a $y$-average,
\be
\bld{\overline{u}}(x)\equiv \int_0^{L_y}\bld{u}(x,y)dy/L_y \ .
\ee
Since $\overline{u}_x=0$ (because $u_x=-\partial_y\nabla^{-2}\omega$), 
the total $x$-momentum always vanishes, and need 
not be considered further. Henceforth we call $M$ the momentum.
From $\bar{\omega}=\partial_x\bar{u}_y$,
we may express
 \begin{eqnarray}
  \overline{u}_y-\overline{u}_y\vert_{\xi^\pm=0}&=&\int_{-\infty}^x(\overline{\omega}(x')-\overline{\omega}(x')\vert_{\xi^\pm=0})dx'  
  \label{eq:ubareq}
  \\
  &=&\mu\left(|\xi_{k_y}^-|^2\delta(x+s)-|\xi_{k_y}^+|^2\delta(x-s) \right)
  \label{eq:ubareq2}
\end{eqnarray}
where the latter relation applies for a single $k_y$ component of $\xi^\pm$ (including the
$-k_y$ piece).
To derive it, note that
the integrand in equation (\ref{eq:ubareq})
 vanishes everywhere except where a line of constant $x'$ pierces
either of the two contours.  Considering first the vicinity of the $\xi^+$ contour, we have
$\overline{\omega}(x')-\overline{\omega}(x')\vert_{\xi^\pm=0}=-\mu|\xi_{k_y}^+|^2d\delta/dx|_{x'-s}$,
after writing $\omega$ in terms of Heaviside's function as below equation (\ref{eq:xi1}), and
then expanding.
The contour $\xi^-$ contributes the analogous amount, but with the opposite sign because the
jump in vorticity across it is in the opposite sense. Therefore both steps contribute
\beqn
M&=&M_-+M_+ \\
M_\pm&\equiv & \mp\mu |\xi_{k_y}^\pm|^2 \label{eq:mompm} \ .
\eeqn
The overall signs of $M_-$ and $M_+$ will play an important role in what follows.
They are determined solely by the sign of the jump in vorticity across the corresponding
step,  independent of the functions $\xi^-$ and $\xi^+$.
We may understand this is as follows. To be definite, we consider first the step at $x=-s$ and
set $\mu>0$.  Because $\bar{\omega}=d\bar{u}_y/dx$, the unperturbed 
$\bar{u}_y$ is constant for $x<-s$ and has positive slope ($=\mu$) for $x>-s$.  
Any perturbation  $\xi^-$ must widen the range in $x$ over which 
$\bar{\omega}$ makes a transition from $0$ to $\mu$.  Hence it will widen
the range over which $\bar{u}_y$ increases in slope.  Therefore the perturbed
$\bar{u}_y$ must exceed its unperturbed value and $M_->0$.  Generalizing this
reasoning, it is clear that 
wherever the unperturbed profile has a positive jump in vorticity, perturbations
always contribute positive momentum; similarly, perturbed negative jumps contribute negative momentum.

The perturbed kinetic energy is
\beqn
E&\equiv& {1\over 2}\int 
\left(
-qx\bld{\hat{y}}+\bld{u}
\right)^2 -(-qx\bld{\hat{y}}+\bld{u})^2\vert_{\xi^\pm=0}
dxdy
\\
&=&
E_-+E_++E_{u^2}
\eeqn
where
\beqn
 E_-+E_+&\equiv& \int-qx(\bar{u}_y-\bar{u}_y\vert_{\xi^\pm=0})dx
 \\
&=&qs\mu \left(|\xi_{k_y}^-|^2+|\xi_{k_y}^+|^2  \right) \ ,
\eeqn
which follows from equation (\ref{eq:ubareq2}),
and 
\beqn
E_{u^2}&\equiv&\int{1\over 2}(\overline{u^2}-\overline{u^2}\vert_{\xi^\pm=0})dx
\\
&=&{\mu^2\over 2|k_y|}\left|
\xi_{k_y}^- -\xi_{k_y}^+
\right|^2 \ ,
\eeqn
which follows from $u_{y,k_y}=(i/k_y)du_{x,k_y}/dx$ and
equation (\ref{eq:xi1}) (including now the exponential previously dropped
from equation (\ref{eq:uxky1})).
There are two different kinds of energy.  The first, composed of $E_-$ and $E_+$,
is localized near each step in $\omega$, and is due to changes in 
$\bar{u}_y$ that are second order in $\xi^\pm$; we call it the mean-flow energy.
The second, $E_{u^2}\geq 0$, is spread out in $x$ over a range $\sim 1/|k_y|\gg s$,
and is due to the terms in $\bld{u}$ that are first order in $\xi^\pm$; we call it
the perturbation energy.

 The equations for $\xi^\pm$ conserve momentum,
\beqn
{dM_\pm\over dt}=\pm F \label{eq:feq}
\eeqn
where
\be
F\equiv {\rm sgn}(k_y){i\over 2}\mu^2\left(\xi_{k_y}^{+*}\xi_{k_y}^--\xi_{k_y^+}\xi_{k_y}^{-*}  \right) \ ,
\ee
is  
 the rate at which momentum flows from the step at $x=-s$ to the one at $+s$.
In terms of more familiar variables, it can be shown that the $x$-dependent 
momentum flux $\overline{u_xu_y} $ is
\be
\overline{u_xu_y} 
  =
 \left\{
 \begin{array}
 {r@{\quad,\quad}l}
 F & |x|<s\\
 0 & |x|>s
 \end{array}  
 \right.  \ ,
\ee
which is in accord with equation (\ref{eq:feq}).
Similarly, the equations for $\xi^\pm$ conserve energy,
\beqn
{d\over dt}(E_-+E_+)
&=&-q(2s)F 
\label{eq:momo}
\\
&=&
-{dE_{u^2}\over dt} \ ,
\eeqn
which may be understood as follows.
Consider two streamlines of the unperturbed flow, 
one at $x=x_1$ and the other at $x=x_2$, with $x_1<x_2$, and label their unperturbed $y$-velocities
$v_1$ and $v_2$.  Their unperturbed specific energies (energy per unit mass) are then
$v_1^2/2$ and $v_2^2/2$.  
We perturb the streamlines by transferring 
$y$-momentum from one streamline to the other.
If the specific momentum transferred from  $x_1$ 
to  $x_2$
is $\Delta v$, then the change in mean-flow energy is
\beqn
&&{(v_1-\Delta v)^2-v_1^2\over 2 }
+{(v_2+\Delta v)^2-v_2^2\over 2 } \nonumber \\
&\approx& (v_2-v_1)\Delta v
= -q(x_2-x_1)\Delta v \label{eq:momtz} \ , 
\eeqn
 which 
 is equivalent
to equation (\ref{eq:momo}).  
When the momentum transfer tends to lower the relative velocity of the two streamlines
(i.e., $\Delta v>0$), 
energy 
is extracted from the mean flow.
Conversely, when the transfer tends to increase the relative velocity, energy must be added to the mean flow.\footnote{Note that outward transport of angular momentum in an accretion disk is equivalent
to a momentum transfer in a shear flow that tends to lower the relative velocity of two streamlines.}
Of course, the total energy is conserved: if energy is extracted from the mean flow it must correspondingly
increase $E_{u^2}$.

Let us see what the above results imply for the shear instability, first setting $\mu<0$
so that the instability can operate.  Perturbations to the step at $x=-s$ have negative momentum,
and those at $x=s$ have positive momentum (eq. [\ref{eq:mompm}]).  Therefore perturbations
transfer momentum from $x=-s$ to $x=s$.  Since this transfer tends to lower the relative velocity
of the streamlines at $x=\pm s$, energy is extracted from the mean flow and added to 
$E_{u^2}$.  This is the recipe for instability.  However, for instability to actually occur 
the increase in $E_{u^2}$ must balance the decrease in mean-flow energy.
Since $E_{u^2}\propto 1/|k_y|$, this is only possible for sufficiently small $|k_y|$.  
Quantitatively, unstable 
perturbations must have $M_-+M_+=0$; otherwise, $M_-+M_+$ could
not remain constant as $|\xi_{k_y}^\pm|$ grows.  Therefore, $|\xi_{k_y}^-|=|\xi_{k_y}^+|$.
Similarly, energy conservation implies $E_{u^2}+E_-+E_+=0$, or
\be
|k_y|={-\mu\over 2qs}(1-\cos \phi) \ ,
\label{eq:kystabz}
\ee
where $\phi$ is the phase difference  between $\xi^+_{k_y}$
and $\xi^-_{k_y}$; this relation also follows from equations (\ref{eq:efngood})-(\ref{eq:gammagood}).
The largest $|k_y|$ 
occurs when $\phi=\pi$, and is equal to
$\kappa=|\mu|/qs$, as for equation (\ref{eq:kapkap}).

If $\mu>0$, equation (\ref{eq:kystabz}) can never be satisfied, and
there is always stability, as implied by Fj\o rtoft's theorem \citep{DR04}. This seemingly
mysterious behavior can also be  understood from momentum and energy considerations. 
When $\mu>0$, the perturbed momentum at $x=-s$ is positive, and that at $x=s$ is negative.  Therefore
any perturbation must transfer momentum from $x=s$ to $x=-s$, which enhances the background shear, and
must increase the mean flow energy $E_-+E_+$.  By energy conservation this is only possible if the
perturbation energy $E_{u^2}$ decreases, 
and since $E_{u^2}$ is always positive, no perturbation can grow.

\subsection{Numerical Simulation of the Top-Hat}
\label{sec:twonum}

Figure \ref{fig:nonlin} shows the nonlinear evolution when the unperturbed vorticity is given by
equation (\ref{eq:tophat}).    The initial perturbation to the curves at $x=\pm s$ is unstable ($k_y=0.75\kappa$), and wraps up
into a  vortex.  

To make this figure we wrote a Lagrangian code, which can handle the rapid
variations in $\omega$ more easily than can the pseudospectral code.
The Lagrangian form of the equation of motion (eq. [\ref{eq:omeq}])
is  $\partial_t \bld{x}=-qx\bld{\hat{y}}+\bld{u}$, where $\bld{x}=\bld{x}(\bld{x_0},t)$ is now a function 
of the initial  coordinate and time, $\partial_t$ is taken at fixed $\bld{x_0}$, and 
$\bld{u}$ is determined by equation (\ref{eq:uinv}).  We chose 2000 initial points along each of
the two curves at $x=\pm s$, and evolved these points assuming periodic boundary conditions
in $y$ and open boundary conditions in $x$. Given 
$\bld{x}$ for each of these points, we immediately know $\omega$, since $\omega$ is locally conserved.
To convert $\omega$ to $\bld{u}$, we evaluated equation (\ref{eq:uinv}) by interpolating
$\omega$ onto a grid, taking the Fourier
transform of $\omega$, multiplying by the appropriate factors of $\bld{k}$,  taking the inverse
Fourier transform to yield $\bld{u}$ on the grid, and finally interpolating from the grid back
onto the curves.
To check the code, we plot in Figure \ref{fig:nonlin} at time$=3000$ the eigenfunction 
\be
 \xi^\pm(y)\propto \pm \cos\left(|k_y| y\mp {\rm tan}^{-1}
{\gamma\over qs|k_y|}\right)
\ee
(eq. [\ref{eq:efngood}]).
These curves are indistinguishable from the code's output, implying agreement between
code and theory.  We have also checked that the growth rate at early times, while $\xi^\pm\lesssim s$, 
is $\xi^\pm\propto \exp(\gamma t)$,
where $\gamma=1/1200$ (eq. [\ref{eq:gammagood}]). As usual, for our numerical results we measure
time in units of $\Omega_0^{-1}$, i.e., we set $q=3/2$.

At late times,  $\omega$ exhibits rapid spatial variations, and the boundaries that separate
the regions where $\omega=0$ from the ones where $\omega=\mu$ become more and
more convoluted.  The Lagrangian code has no viscosity; the evolution shown in Figure \ref{fig:nonlin}
breaks down at time $\gtrsim 10,000$ when the curves are so convoluted that adjacent points
on a curve are widely separated.
If we
were to include an explicit viscosity, it would wipe out the rapid variations in $\omega$ at late times, 
leaving a smooth vortex.  At what stage the viscosity acts depends on how small the viscosity is.  In astrophysical disks, the viscosity is extremely small, so if a shear instability acts, it might lead
to extremely rapid variations in $\omega$.  Whether this is unstable to three dimensional perturbations
is an interesting possibility.

\includegraphics*[width=9cm]{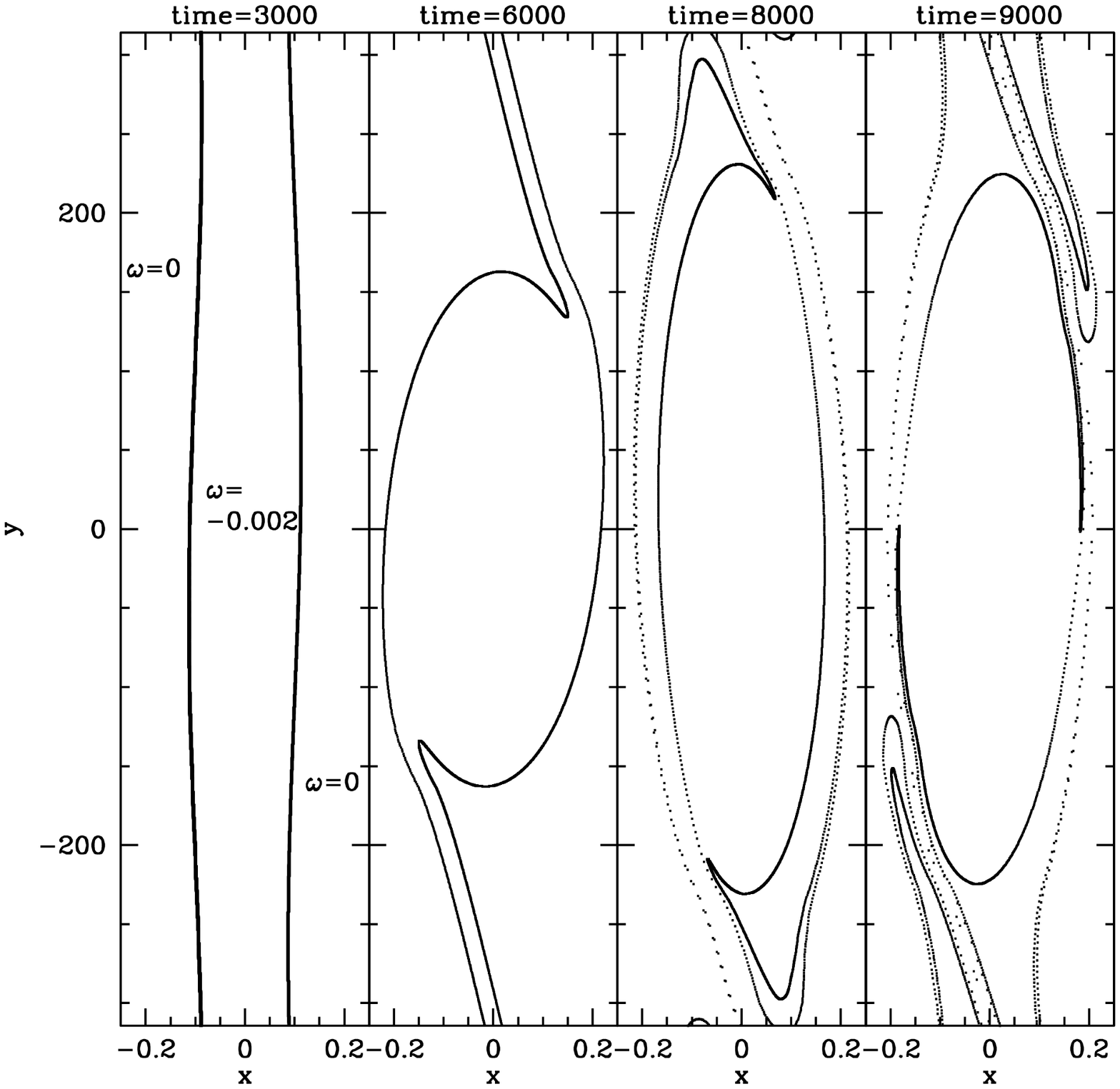}
\figcaption{
{\bf Numerical Simulation of the Top-Hat:
The unperturbed vorticity is given by equation (\ref{eq:tophat}),
with $s=0.1$ and $\mu=-0.002$; the initial perturbation to the curves
at $x=\pm s$, $\xi^\pm(y)$, are sinusoidal with $k_y=0.01=0.75\kappa$, and 
amplitudes $=0.001$.
The initial perturbation is unstable to the shear instability, and wraps up into a  vortex.
The numerical simulation was done with a Lagrangian code described in the main text.
\label{fig:nonlin}
}}

\subsection{Infinite Number of Steps}
\label{subsec:infinite}

Consider now the general case, in which the   unperturbed $\omega$ is given by
$\omega_{\rm unp} = \omega^{i}  \ , \ {\rm for\ }x_i<x<x_{i+1} \ ,$
where   $\omega^{i}$ and $x_i$ are sets of constants with $i=1,\cdots,\infty$.
As before, there is a curve $\xi^i(y,t)$  at
each step $x_i$, whose equation of motion 
is, generalizing equation (\ref{eq:xikyeq}),
\be
\left({d\over dt}
-qx_iik_y\right)
\xi_{k_y}^{i}=u_{x,k_y} \label{eq:fte} \ .
\ee
Equation (\ref{eq:xi1}) generalizes to
\be
u_{x,k_y} = -{i\over 2}{\rm sgn}(k_y)\sum_i \omega_\delta^i \xi_{k_y}^{i} \ .
\label{eq:uxky}
\ee
where the jump in vorticity at step $i$ is
\be
\omega_\delta^i\equiv \omega^i-\omega^{i-1}
\ee
We continue to assume that $|k_y|s\ll 1$, where $s$ is now the extent of the region
of non-zero vorticity.
To find instability, we set
 $\xi_{k_y}^i\propto e^{\gamma t}$, 
yielding for the eigenfunction and eigenvalue 
\beqn
\xi_{k_y}^i
&\propto& {1\over \gamma-qx_iik_y} \label{eq:efn2}
\\
1 &=& -{i\over 2}{\rm sgn}(k_y)\sum_i {\omega_\delta^i\over \gamma-qx_iik_y} \ . \label{eq:drlang}
\eeqn
thus recovering the results derived with swinging waves 
(eqs. [\ref{eq:xi}],[\ref{eq:dr2}]) when we make the identification $\gamma=\alpha q \bar{K}_y$.

\subsection{Top-Hat With Wings}
\label{sec:sunk}

Because the top-hat vorticity profile has $\int\omega dxdy\ne 0$,
it produces a velocity field $u_y$ that is infinite in extent.  Although this is somewhat
disconcerting, profiles that do not suffer from this problem behave similarly to the top-hat.
To illustrate, we consider the top-hat with wings, i.e.,
\be
 \omega_{\rm unp}
  =
 \left\{
 \begin{array}
 {r@{\quad,\quad}l}
 \mu & |x|<s \\
 -\mu & s<|x|<2s \\
 0 & 2s<|x|
 \end{array}  
 \right. 
 \ , \label{eq:wings}
\ee
which clearly integrates to zero.
Figure \ref{fig:wings} shows a numerical simulation of this profile, done with the Lagrangian
code described in \S \ref{sec:twonum}.  See caption for details.  For unstable initial conditions, 
the vorticity again tends to wrap up into a vortex.

It is perhaps not surprising that the infinite extent of the unperturbed $u_y$ 
is not important, since the unperturbed $u_y$ was not important for any of our three
explanations of the shear instability, in \S\S \ref{sec:ssw}, \ref{sec:tophat}, and \ref{sec:momen}.

\includegraphics*[width=9cm]{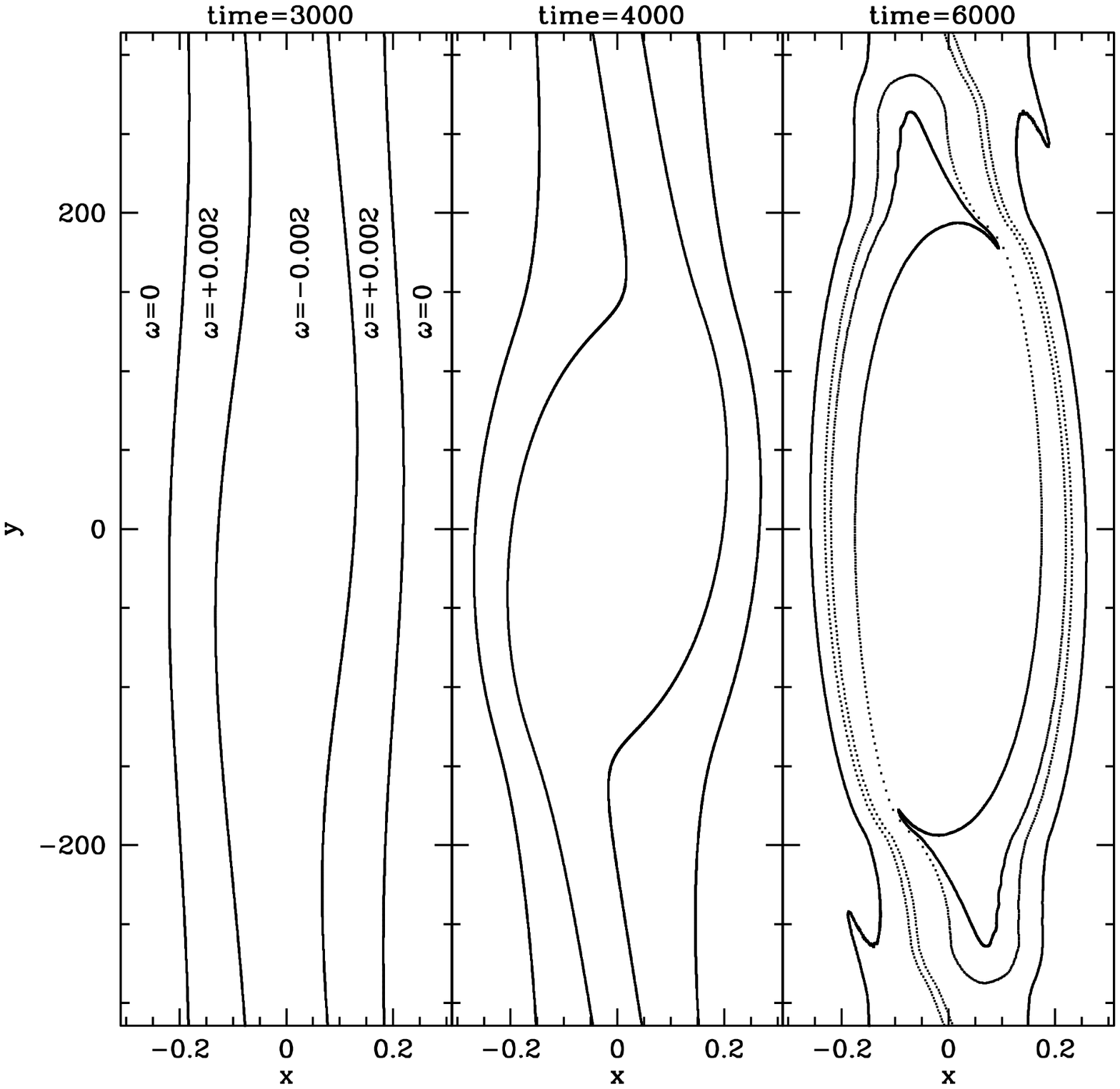}
\figcaption{
{\bf Numerical Simulation of Top-Hat with Wings:
The unperturbed vorticity is given by equation (\ref{eq:wings}),
with $s=0.1$ and $\mu=-0.002$; the initial perturbations
 are sinusoidal with $k_y=0.01$.
 The dispersion relation (eq. [\ref{eq:drlang}]) shows that this situation is unstable, with
 growth rate $\gamma=1/810$.  We checked that at early times the numerical simulation
 reproduces this growth rate.
 As in Figure \ref{fig:nonlin}, the nonlinear outcome is a tightly wrapped vortex with convoluted boundaries.
\label{fig:wings}
}}

\section{Conclusion}
\label{sec:summary}

Two dimensional shear flows are linearly stable, but nonlinearly unstable.
Linear modes that are aligned with the shear (``axisymmetric'') have zero frequency;
those that are not aligned (``swinging'') are transiently amplified before decaying.
When the two couple nonlinearly, they are  
always unstable
provided that the swinging mode's azimuthal wavelength is long enough.
If this is the case, the swinging mode couples with the axisymmetric one to produce
a new leading swinging mode with larger amplitude than itself, implying instability.
Quantitatively, the criterion for instability is
\be
{k_{y,{\rm sw}}\over k_{x,{\rm axi}}} \lesssim {|\omega|\over q} =
{|\omega|\over 3\Omega/2}
\ , \label{eq:instab}
\ee
where $|\omega|/q$ is the ratio of perturbed to background vorticity, and
${k_{y,{\rm sw}}/ k_{x,{\rm axi}}}$ is the ratio of swinging-azimuthal to 
radial-axisymmetric wavenumbers.  This shows that even
 when the amplitudes of the perturbations
are vanishingly small, instability results for small enough $k_{y,{\rm sw}}$.

When the instability is triggered, the swinging waves grow exponentially at the rate 
$\sim 1/|\omega|$, assuming that their wavelength is larger than the critical value
(eq. [\ref{eq:instab}]) by a factor of order unity.   Eventually, regions of negative 
vorticity concentrate into vortices with increasingly convoluted boundaries.  
Since
astrophysical disks have very little viscosity, such vortices might become turbulent
in three dimensions before viscosity can act to smooth them.

In \S \ref{sec:piece} we explained how 
the instability works in real space.   When a nearly axisymmetric strip 
of perturbed vorticity has a small bulge, this bulge tends
to tilt the strip.  Advection of the tilted strip by the background shear reinforces the bulge
when equation (\ref{eq:instab}) holds, where $k_{x,{\rm axi}}^{-1}$ is now interpreted
as the width of the strip, and $k_{y,{\rm sw}}$ is the azimuthal wavenumber
of a perturbation to the strip. (It is also required that $\omega<0$ if the strip only has a single sign
of vorticity.)
We also  showed that the instability can be understood from momentum and energy considerations:
when equation (\ref{eq:instab}) holds and the strip is unstable,  the momentum transferred from a streamline
on one side of the strip to the
other
 tends to lower
 the relative velocity of two streamlines (which is equivalent to transferring  angular momentum outwards in 
an accretion disk), and therefore converts shear energy into
perturbation energy.

It is tempting to speculate that this nonlinear instability might provide a route to turbulence.  
Perhaps vortices are unstable to three dimensional perturbations, and these perturbations
can  create new axisymmetric modes that are in turn unstable to the production of vortices. 
It is also possible that other effects are required.  Perhaps vorticity is produced
by convection or by stirring from planets, and if enough vorticity is produced, 
the disk becomes unstable in the manner described in this paper.
Even if other effects are required,
we feel  that the technique introduced in this paper---the examination
of couplings between swinging modes---will be helpful.  It gives a well-defined procedure for analyzing
 nonlinear effects, both theoretically and numerically.
Similar techniques might also be employed to give insight into
other issues in
 shear flows, such as the nonlinear
development of the magnetorotational instability.

\acknowledgments{We thank Gordon Ogilvie for showing us how easy it is to implement
shearing box boundary conditions in  a pseudospectral code, and
 Jeremy Goodman for helpful discussions
at an early stage of this project. We also thank the referee for helpful suggestions.}

\appendix
\section{Description of the Pseudospectral Code}

The pseudospectral code simulates the 3-D shearing box equations of motion 
(eqs. [\ref{eq:eom2}]-[\ref{eq:eom1}]), though in the present paper we only present 2-D simulations. 
Most of our methods are standard, except for our new method for remapping
modes that is both simpler and more accurate than that used by other investigators.

In our simulations, the unit of time is chosen so that
\be
\Omega_0=1 \ , 
\ee
and we also set the constant density
\be
\rho=1 \ .
\ee
Equations (\ref{eq:eom2})-(\ref{eq:eom1}) are solved in a box of size $L_x\times L_y\times L_z$ that is subject to
periodic boundary conditions in $y$ and $z$,
\begin{eqnarray}
\bld{u}(x,0,z)&=&\bld{u}(x,L_y,z) \\
\bld{u}(x,y,0)&=&\bld{u}(x,y,L_z)
\end{eqnarray}
and to a shifted periodic (``shearing sheet'')
boundary condition in $x$:
\begin{equation}
\bld{u}(0,y,z)=\bld{u}\left(L_x,\left(y-qtL_x\right){\rm mod} L_y,z\right) \ . \label{eq:shifted}
\end{equation}
The modulus function ``mod'' is defined as follows: 
\be
(y-qtL_x)\ {\rm mod}\ L_y\equiv y-qtL_x + pL_y \label{eq:mod}
\ee
where $p$ is the integer that makes the right-hand side of the above expression lie within
the box in the $y$-dimension:
$0\leq y-qtL_x+pL_y<L_y$.

Changing to Lagrangian coordinates $\bld{X},T$ that shear with the background velocity $-qx\bld{\hat{y}}$
(eqs. [\ref{eq:y}]-[\ref{eq:xt}]),
equation (\ref{eq:eom2}) becomes
\be
{\partial\bld{u}\over\partial T}=
2u_y\bld{\hat{x}}-(2-q)u_x\bld{\hat{y}}-\bld{\nabla_x}P-\bld{u\cdot\nabla_x u}
\label{eq:eom2a}
 \ ,
\ee
where $\bld{\nabla_x}\equiv\bld{\nabla}$  is a gradient with respect to $\bld{x}$, not $\bld{X}$.

Lagrangian coordinates
 $(X,Y,Z)$ lie within a box of size $L_x\times L_y\times L_z$; the boundary conditions
on the sides of this box are periodic in all three dimensions.
The Fourier transform of $\bld{u}$ with respect to $\bld{X}$ is
\be
\bld{{u}}_{J_x,J_y,J_z}\equiv {1\over L_xL_yL_z}\int_{L_x\times L_y\times L_z}
 \!\!\!\!\!\!\!\!\!\!\!\!\!\!\!\!\!\!\! 
\bld{u}
\exp{\left(- i \bld{K\cdot X}
\right)}
d^3\bld{X} \ ,  \label{eq:fftcont}
\ee
where the wavevector
$\bld{K}$ takes on the discrete values
\be
K_x={2\pi\over L_x}J_x \ \ , \ K_y={2\pi\over L_y}J_y \ \ ,\  K_z={2\pi \over L_z}J_z \ , \label{eq:bigk}
\ee
with $J_x,J_y,J_z$  integers. Hence the Fourier transform of equation (\ref{eq:eom2a}) with respect to $\bld{X}$ is 
\be
{d\bld{\tilde{u}}\over d T} = 2\tilde{u}_y\bld{\hat{x}}-(2-q)\tilde{u}_x\bld{\hat{y}}-i\bld{{k}}(\bld{K},T)\tilde{P}-i\bld{{k}}(\bld{K},T){\bld\cdot}\widetilde{[\bld{uu}]} \ ,  \label{eq:eom2b}
\ee
where, for clarity, we 
denote Fourier transforms with a tilde instead of the subscripts $J_x,J_y,J_z$.
 The $(x,y,z)$ components of the wavevector $\bld{k}(\bld{K},T)$ are defined as follows
\be
\bld{k}(\bld{K},T)\equiv
 (K_x+qTK_y,K_y,K_z)  \ .
\label{eq:kx}
\ee
The last term in equation (\ref{eq:eom2b}) has been written in a tensor form;
 it represents a vector whose $j'$th component is 
\be
-i \sum_{l=1}^3 k_l\widetilde{u_lu_j} \ , \label{eq:noabs}
\ee
where $\widetilde{u_lu_j}$ is the Fourier transform of the product $u_l u_j$.
Equation (\ref{eq:eom2b}) is supplemented by the Fourier-space version of equation (\ref{eq:eom1}):
\be
\bld{k}(\bld{K},T)\bld{\cdot}{\bld{\tilde{u}}}=0 \label{eq:eom1b}
\ee

\label{sec:numerical}

The numerical code
 evolves equation (\ref{eq:eom2b}) in time using second-order Runge-Kutta, with $\tilde{P}$ determined by the incompressibility constraint (eq. [\ref{eq:eom1b}]).  Equation (\ref{eq:eom2b}) is integrated for $n_x\times n_y\times n_z$ different modes, each of which is labelled by the integers $(J_x,J_y,J_z)$, with
$-n_x/2<J_x\leq n_x/2$, $-n_y/2<J_y\leq n_y/2$, and $-n_z/2<J_z\leq n_z/2$.\footnote{In truth, 
only half of these modes are simulated, which makes the code run twice as fast as it otherwise would.
The other half that are not simulated are determined by the condition
 that the components of $\bld{u}$ are real,
 not complex, numbers.
But  for the purposes of the discussion in the present section, we ignore this complication.}

The code equations differ from  the exact equations (\ref{eq:fftcont})-(\ref{eq:eom1b}) in two ways.
First, 
the Fourier transform in the code is  the discretized version of equation (\ref{eq:fftcont}):
\be
\bld{{u}}_{J_x,J_y,J_z}\equiv {1\over n_xn_yn_z} \sum \bld{u}(\bld{X})\exp\left(
- i \bld{K\cdot X} 
\right) \ , 
\label{eq:fftdisc}
\ee
with the sum taken over $n_x\times n_y\times n_z$ discrete Lagrangian points ${\bld{X}}$.
And second, $k_x$ 
is not given $K_x+qTK_y$, as would be inferred from equation (\ref{eq:kx}).
  In this respect, our code differs from other codes
\citep[e.g., ][]{Rogallo81,UR04,BM06}.
A single mode with a given $(J_x,J_y,J_z)$ has the following spatial dependence
\beqn
\exp 
{\left(2\pi i\left(
{J_xX\over L_x}+{J_yY\over L_y}+{J_zZ\over L_z}\right)\right)} \\
=\exp{\left(
2\pi i
\left(
\left({J_x\over L_x}+qT{J_y\over L_y}\right)x+{J_y\over L_y}y+{J_z\over L_z}z
\right)
\right)} \ .
\eeqn
In substituting for $Y$ with equation (\ref{eq:y}), the modulus term may be dropped
because it does not contribute to the exponential.  By the same token, we can replace $J_x$ on the right-hand side of the above expression with $J_x+pn_x$, where $p$ is any integer, without affecting the exponential.  (Note that  $x=\{0,1,...,n_x-1\}L_x/n_x$.)  This is the well-known phenomenon of aliasing. So instead of equation (\ref{eq:kx}) for $k_x$, we can equally well use a similar expression, but with $J_x\rightarrow
J_x+pn_x$.  
The value of $p$ should be chosen so that the physical wavenumber $k_x$ is shifted as close as possible to zero.  Other $p$'s correspond to lengthscales smaller than the grid's scale.
Hence,
\be
k_x\equiv {2\pi\over L_x}\left(\left(J_x+qTJ_y{L_x\over L_y} \right)\widehat{{\rm mod\ }} n_x\right) \ , \label{eq:goodkx}
\ee
where $\widehat{\rm mod}$ is a modulus operator: 
\be
J\ \widehat{\rm mod} \ n_x\equiv J+pn_x \ ,
\ee
with $p$ an integer such that $-n_x/2<J+pn_x\leq n_x/2$.  (These limits differ
 from those used to define the earlier modulus operator below eq. [\ref{eq:mod}].)
Since other authors use   $k_x=K_x+qTK_y$ instead of
equation (\ref{eq:goodkx}) for $k_x$, 
 the $k_x$ of each of their modes grows
 to large positive or negative values.  To stop this growth, they periodically remap 
 the modes to small $k_x$.  But this has the undesirable consequence that the distribution of $k_x$'s that  is being used to simulate the fluid changes substantially from just after a remapping event to just before the subsequent remapping.  Conversely, with equation (\ref{eq:goodkx}) there is no need for an explicit  remapping; in effect, the remapping is done automatically by the modulus operator.

The nonlinear term in equation (\ref{eq:eom2b}),  written explicitly in equation (\ref{eq:noabs}), is evaluated using the standard pseudospectral method \citep[e.g., ][and references therein]{MG01}: 
after dealiasing (see below), $\hu$ is transformed into $\bld{u}$ by performing an inverse fast Fourier
transform (FFT).
Then, the product $u_iu_j$ is formed in real space, and finally the FFT of this product is taken.

To prevent aliasing errors when evaluating the nonlinear term, $\hu$ is truncated just before
the time derivative is evaluated by setting to zero any mode whose 
 $\vert k_x\vert$, $\vert k_y\vert$, or $\vert k_z\vert$ exceeds 2/3 of its maximum value---the
standard ``2/3-rule'';   explicitly,  $\bld{{u}}_{J_x,J_y,J_z}\rightarrow 0$ wherever one of the following hold
\begin{eqnarray}
\vert J_y\vert&>&{n_y\over 3}  \\ 
 \vert J_z\vert&>&{n_z\over 3}  \\
  \left\vert
\left(J_x+qTJ_y{L_x\over L_y}\right)\widehat{\rm mod}\  n_x
\right\vert& >& {n_x\over 3} \label{eq:dealias}
\end{eqnarray}
The last condition above follows from  equation (\ref{eq:goodkx}).
Note that the $k_x$ of a mode is remapped by the modulus operator from $+k_x$ to $-k_x$
when $  \left\vert
\left(J_x+qTJ_y{L_x/ L_y}\right)\widehat{\rm mod}\  n_x
\right\vert ={n_x/ 2}$.
Such modes have had their amplitude set to zero because they satisfy the inequality of equation (\ref{eq:dealias}).  Hence there is no danger that remapping
artificially introduces power into leading modes.

\end{document}